\documentclass[12pt,preprint]{aastex}
\usepackage{epsfig,amssymb}
\shorttitle{Acceleration in Evolving Networks}
\shortauthors{Vlahos et al.}

\received{2003 November 20}
\begin{document}


\title{Particle Acceleration in an Evolving Network of Unstable Current Sheets}


\author{Loukas Vlahos, Heinz Isliker and Fabio Lepreti}
\affil{Department of Physics,\\ University of Thessaloniki,
Thessaloniki 54 124, Greece}



\begin{abstract}
We study the acceleration of electrons and protons interacting
with localized, multiple, small-scale dissipation regions inside an evolving,
turbulent active region. The dissipation regions are 
Unstable Current Sheets (UCS), and in their ensemble they form a complex, 
fractal, evolving network of acceleration centers. 
Acceleration and energy dissipation are thus assumed to be fragmented.
A large-scale magnetic topology
provides the connectivity between the UCS and determines in this way the 
degree of possible multiple acceleration. The particles travel along the
magnetic field freely without loosing or gaining energy, till they
reach a UCS. In a UCS, a variety of acceleration
mechanisms are active, with the end-result that the
particles depart with a new momentum.
The stochastic acceleration process
is represented in the form of Continuous Time Random Walk (CTRW),
which allows to estimate the evolution of the
energy distribution of the particles. It is found that under certain conditions
electrons are heated and accelerated to energies above $1\,$MeV 
in much less than a second.
Hard X-ray (HXR) and microwave spectra are calculated from the electrons'
energy distributions, and they are found to be compatible with 
the observations. Ions (protons) are also heated and accelerated, 
reaching energies up to 10 MeV almost simultaneously with the electrons.
The diffusion of the particles inside the active
region is extremely fast (anomalous super-diffusion).
Although our approach does not provide insight into the
details of the specific acceleration mechanisms involved, 
its benefits are that it relates acceleration to the energy 
release, it well describes the stochastic nature of the acceleration
process, and it can incorporate the flaring large-scale magnetic
topology, potentially even its temporal evolution.
\end{abstract}




\section{Introduction\label{Sec1}}

Solar flares remain, after almost one hundred years of intense
study, an unsolved problem for astrophysics. We define as {\it flare}
the sporadic transformation of magnetic energy to (1) plasma
heating, (2) particle acceleration, and (3) plasma flows. It seems
that the total magnetic energy released in a single flare is not
equally spread to all three components. The energetic particles
carry a very large fraction of the total energy released during a
flare, reaching sometimes up to $50\%$ \citep{Saint02}.

Modeling the explosive energy release requires methods that can
treat simultaneously the large scale magnetic field structures and
the small-scale dissipation events. 
The convection zone actively participates in the formation and
evolution of large scale structures by rearranging the position of
the field lines and at the same time it adds new magnetic flux
(emerging flux) and new stresses to the existing topologies. The
loss of stability of several loop-like structures forms large
scale disturbances (CME), which further disturb the pre-existing
and so-far stable large scale structures. In this way, the 3-D
magnetic topologies are constantly forced away from the potential
state (if they ever reach one) due to slow or abrupt changes in
the convection zone. All these non-potential magnetic stresses
force the large-scale magnetic topology to form short-lived, 
small-scale magnetic discontinuities in order to dissipate the excess energy
in localized current sheets. This conjecture was initially
proposed by \citet{Parker88}, and subsequently it was modeled
using 3-D MHD numerical simulations by many researchers  
\citep[see for example the work of][]{NorGal96}. \emph{We emphasize here
that the concept of sudden formation of a distribution of unstable
discontinuities inside a well organized large-scale topology has
not been appreciated or used extensively enough to model the solar
flare phenomenon}.

The scenario of spatially distributed, localized, small-scale dissipation,
which moreover evolves in time, is not least supported by the observations
which indicate highly fragmented energy dissipation and particle 
acceleration processes. There is strong evidence that narrow-band 
milli-second spike-emission in the radio range is directly associated to 
the primary energy release events. The emission itself of the radio-spikes 
is fragmented in space and time, as is seen in radio-spectrograms and in 
spatially resolved observations. It must thus be concluded that also the 
energy release process is fragmented in space and time, to at least
the same degree as is the radio spike-emission \citep[see][]{Benz03}.
Also type III burst radio-emission, caused by electron beams escaping from 
flaring regions, exhibits fragmentation as a strong characteristic  
\citep[e.g.][]{Benz94}.
It is a too simple interpretation of the available data that the large 
scale structures seen have a relatively simple topology down to all scales.

One approach which is capable to capture the full extent of this
interplay of highly localized dissipation in a well-behaved large
scale topology ('sporadic flaring') is based on a special class of
models which use the concept of Self-Organized Criticality 
\citep[SOC;][]{Bak87}.
The main idea is that active regions evolve by the continuous
addition of new or the change of existing magnetic flux on an existing 
large scale
magnetic topology, until at some point(s) inside the structure
magnetic discontinuities are formed and the currents associated with them
reach a threshold. This causes a fast rearrangement of the local
magnetic topology and the release of the excess magnetic energy at
the unstable point(s). This rearrangement
may in-turn cause the lack of stability in the neighborhood, and so forth,
leading to the
appearance of flares (avalanches) of all sizes that follow a well defined
statistical law \citep{Lu91,Lu93,Vla95,Isl00,Isl01}, which agrees
remarkably well with the observed flare statistics \citep{cros}.

The modeling of the solar atmosphere cannot be done exclusively
with the use of high-resolution MHD numerical codes.
The small scale discontinuities easily give rise to kinetic
instabilities and anomalous resistivity, which
play a very important role, they dramatically change the
evolution of even the large scale structures.
On the other hand, numerical codes following the
evolution of charged particles in idealized, non-evolving, large-scale
current-sheets also fail to capture the spatio-temporal evolution of
the magnetic energy dissipation.
The extraordinary efficiency of particle acceleration
during solar flares questions the use of
ideal MHD for the description of flares, 
since it misses kinetic plasma effects, which yet
play a major role in the energy dissipation process and
for the local state of the plasma. The coupling of the large
scale and the small scale is extremely difficult to handle, not only
in solar physics but in physics in general, and it is the main
reason for not having resolved the solar flare problem for so many
years \citep{Carg02}.

Our inability to describe properly the coupling between the MHD
evolution and the kinetic plasma aspects of the driven flaring region is
the main reason behind our lack of understanding of the
mechanism(s) which causes the acceleration of high energy
particles. Let us now define more accurately the so called
`acceleration problem'. \textit{We need to understand the
mechanism(s) which accelerate electrons and ions in relatively
large numbers to energies well above the relativistic regime on a
short time scale with specific energy-spectra for the different
isotopes and charge states}.

Let us summarize very briefly the main observational constraints
for the acceleration processes in solar flares.

\textbf{Electron} energies well in excess to 100 keV, and
occasionally up to tens of MeV, are inferred. Electrons reach
$100$keV in $<1$s and higher energies in a few seconds. The
number of electrons required above 20 keV is large for x-class
flares and can reach up to $10^{38}$ electrons per second,
although this number is model dependent and not very accurate
\citep[see e.g.][]{Mill97}. The observed spectra
in the hard X-rays can be fitted with single or double power-laws, 
combined often --- but not always --- with a thermal emission spectrum 
at the lowest energies \citep{Hol03,Hol03b,Piana03}. 
The spectra in the microwaves roughly show a power-law decay 
at high frequencies \citep{Hol03}.

\textbf{Ions} (especially protons) are inferred to have energies
above $1\,$MeV and up to $1\,$GeV per nucleon. The 10 MeV ions are accelerated
on the same time scales as the electrons, with the high energy
ions delayed up to $10\,$s.  The total number of ions accelerated may
carry the same energy as the electrons. The continuous component
of the $\gamma$-ray spectra are
usually power laws. An important finding is the massive
enhancement of $^3_2$He in very impulsive flares \citep[see e.g.][]{Mill97}.

\noindent
\textbf{Acceleration mechanisms:}
Numerous books and reviews have been devoted to the challenging
problem of particle acceleration
\citep{Hey,vla84,Mel,vla94,Kir,vla96,Kuip,Mill97}. The
proposed models usually address parts of the problem, and almost
all acceleration mechanisms have no clear connection to the
large-scale topology and to the magnetic energy release mechanism(s). 
The most prominent mechanisms are shock waves 
\citep{HolPes83,Bla87,Ell85,Dec88}, MHD turbulence \citep{Fermi,Mill95}, 
and DC electric fields \citep{BenHol92,Mo85,Mo90}. 

\noindent
\textbf{Mixing acceleration mechanisms:}
Several inquiries have been made in which different acceleration mechanisms
had been mixed.

\citet{Dec} analyzed the role of Shock Drift Acceleration (SDA) when
the shock is surrounded by a turbulent spectrum. As we already
mentioned, the SDA is fast but not
efficient since the particles drifting along the shock-surface's
electric field quickly leave the shock. The presence of
turbulence reinforces the acceleration process by providing a
magnetic trap around the shock surface and forcing a particle to
return many times to the shock surface. The particle leaves the
shock surface, travels a distance $s_i$ inside the turbulent
magnetic field, returns back to the shock surface with
velocity $v_i$, drifts a distance $l_i$ along the shock
electric field $E_{sc}$, changing its momentum by $\Delta p_i\sim
E_{sc}\cdot (l_i/v_i)$, it escapes again, travels a distance
$s_{i+1}$ before returning back to the shock and drifting along the
electric field, in other words, the acceleration-cycle has begun again. 
The process repeats
itself several times before the particle gains enough energy to
escape from the turbulent trap around the shock surface. Let us
note some very important characteristics of this acceleration: (1)
The distances $s_i$ traveled by the particle before returning back to the
shock are only indirectly relevant to the acceleration, they basically
delay the process and influence the overall timing,
i.e.\ the \textit{acceleration time}, another important parameter of the particle
acceleration process. (2) The energy
gain depends critically on the lengths $l_i$ the particle
drifts along the shock surface, in a statistical sense though,
i.e.\ on the distribution of the $l_i,\,i=1,2,3\,...$.
(3) The times $\tau_i$ a particle spends on the shock surface are
again crucial for the energy gain, and also, together with the
$s_i$, for the estimation of the acceleration time.
(4) For the total acceleration problem, which concerns the energies reached and
the times needed to reach them, all three variables, $s_i,\,l_i,\,\tau_i$,
are of equal importance.

\citet{Amb} discussed a similar problem, placing a current sheet
into a region of Alfv\'enic turbulence. Also here, the ability of
the associated DC electric field to accelerate particles is
enhanced by the presence of the MHD turbulence. The acceleration
process is again of a cyclic nature, as in the case of turbulent
SDA, and the process is again characterized by the three variables
$s_i,\,l_i,\,\tau_i$. The turbulent current sheet has several
avenues to enhance the acceleration-efficiency since the plasma inflow is
dynamically driven and causes a variety of new and still
unexplored phenomena. \citet{Azn} analyzed the mixture of
stationary MHD turbulence with a DC electric field. The trapping
of the particles inside the turbulent magnetic field causes a new
`collision scale', and, in some circumstances, acceleration
becomes dependent on an alternative 'Dreicer field', in which
particle collisions are replaced by collisions with magnetic
irregularities. Actually also the diffusive shock acceleration
described above is of a mixed type, having as elements a shock and
magnetic turbulence, although turbulence plays a more passive role
of just scattering the particles. It seems that most acceleration
mechanisms are more or less of a mixed type.

We can conclude that the mixture of mechanisms enhances the
acceleration-efficiency and removes some of the draw-backs
attached to the different, isolated mechanisms. A second, main
conclusion we draw is that \textit{cyclic processes}, e.g.\
through trapping around the basic accelerators, are important
elements --- if not the presupposition --- of efficient and fast
acceleration in space plasmas.

\noindent
\textbf{Can the UCS naturally provide the unification of all
             the above mechanisms~?}
Unstable Current Sheets (UCS) are the regions where magnetic
energy is dissipated, and it is natural to
ask if they can become the actual source of the energetic particles.
According to the existing understanding of UCSs, several
potential mechanisms for particle acceleration
co-exist at a UCS. Plasma flows driving turbulence, shock
waves, and DC electric fields are expected to appear
simultaneously inside and around a driven and evolving UCS. If the UCS is
located in the middle of a turbulent magnetic topology, all these
phenomena will be enhanced and the sporadic external forcing of
the plasma inflow into the UCS will create bursts of
sporadic acceleration. The scenario of a single UCS currently enjoys very
large popularity \citep[e.g.][]{Litv03,FletMart98,Mart88}, 
and the question is: Can one single UCS be
the answer to the acceleration problem~?

We believe that this is impossible since a single, isolated UCS
must be enormously large $(10^{9}\times 10^{9}\times10^5$ m),
remain stationary for a long time, and continuously accelerate
particles with extreme efficiency in order to provide the required
numbers of accelerated particles and the observed acceleration
times. From the Earth's magnetic tail, it is known that large UCS
break up quickly, creating a network of smaller scale UCS, with a
specific probability distribution $P(l)$ of their characteristic
scales, a process which just is a manifestation of turbulence 
\citep{Angelo99}. 
The formation of a large scale helmet above certain loops, driven by
erupting filaments cannot be excluded entirely and represents a
special class of very energetic phenomena \citep{Masu}. We though
believe that such a current sheet breaks down on a very short time
scale, and the formation of smaller scale current sheets will be
unavoidable. Eventually, even in this very special occasion the
acceleration probably takes place in an environment similar to
the one discussed in this article.

There is a second reason,  based on a result on the statistics of flares, 
for questioning the idea of a single, large-scale UCS that is associated 
with just one or at most two loops, as in the scenario of the sometimes 
called 'standard model' \citep[see e.g.][]{Shibata95}. 
\citet{With00} determined the frequency-distributions in total emitted energy 
of flares occurring in the \textit{same, individual} active region 
(number of flares per unit energy), and they find featureless power-law 
distributions, extending over many decades. If a single UCS would be the
basic mechanism behind a flare, then the energy output must be
expected to be 
related to the physical properties of the individual active
region, e.g.\ to the linear dimensions of the active region and 
the associated UCS, 
so that it is at least difficult to imagine how a single-site reconnection 
model could 
be able to produce a featureless energy distribution that extends over many 
decades.

\noindent
\textbf{From single to multiple, small-scale UCS acceleration:}
Our attempt in this article is to take advantage of the positive properties
of isolated UCS as accelerators, but at the same time to assume
that the dissipation happens at \textit{multiple}, \textit{small-scale}
sites.

The 3-D magnetic topology, driven from the convection zone,
dissipates energy in localized UCS, which are spread inside the
coronal active region, providing a natural fragmentation for the energy
release and a multiple, distributed accelerator.
In this way, the magnetic topology acts as a host for the UCS, and the
spatio-temporal distribution of the latter defines the type of a flare,
its intensity, the degree of energization and acceleration of the
particles, the acceleration time-scales etc.
Evolving large-scale magnetic topologies provide a variety of
opportunities for acceleration which is not restricted to flares,
but can also take place before a flare, and after a
flare, being just the manifestation of a more relaxed, but still driven
topology. Depending thus on the level to which the magnetic
topology is stressed, particles can be accelerated without a flare to
happen, and even long-lived acceleration in non-flaring active regions
must be expected to occur.
Consequently, the starting point of the model to be introduced below 
is a driven 3-D magnetic topology, which defines a time-dependent spatial 
distribution of UCS inside the active region. In this article, 
we will focus on the interaction of the UCS with particles.
The details of the mechanisms involved in
the acceleration of particles inside the UCS are not essential in
a stochastic modeling approach. 

Ideas related to the basic approach of this article have been presented earlier
\citep{Ana91,vla93,Ana94,Ana97}, 
but remained conceptually simpler, they left many points open for future 
development, being not that 
general and complete as approaches as the model we present here. 
In the approach we follow here, we 
make extensive use of new developments in the theory of SOC models 
for flares.
Also taken into account are ideas from the theory of
Complex Evolving Networks \citep{Alb,Don}, adjusted though to the
context of plasma physics:
The spatially distributed, localized UCS can be viewed as a network,
whose 'nodes' are
the UCS themselves, and whose 'edges' are the possible particle trajectories
between the nodes (UCS). The particles are moving around in this network,
forced to follow the edges, and undergoing acceleration when they
pass by a node (see also Fig.\ \ref{plottwo}). The network is complex in that it has a non-trivial
spatial structure, and it is evolving since the nodes (UCS) are short-lived,
as are the connectivity channels, which ever change during the evolution
of a flare. This instantaneous connectivity of the UCS
is an important parameter in our model, it determines to what degree 
multiple acceleration is imposed onto the system, which in turn influences the
instantaneous level of energization and the acceleration time-scale of the 
particles.

In this article, we introduce a stochastic, multiple acceleration
model for solar flare particle acceleration. We assume an ensemble
of unstable current sheets (UCS), distributed in space in such a way
that it reflects a relatively simple large scale topology in which
turbulence at small scales is developing. Concretely, the
UCS in their ensemble form a fractal, as it is the case in the
SOC models \citep{Isl03b,Mc02}. Particles move erratically around, 
and occasionally
enter a UCS in which they undergo acceleration. The acceleration
process is taken as a simple DC electric field mechanism.
The basic approach of the model is a combined random walk in position-
and momentum-space. The frame-work we introduce is such that
potentially all multiple acceleration models can be formulated
in this frame-work, as long as the acceleration takes place in
spatially localized, disconnected, small-scale regions.
Our main interest here is in the diffusion, acceleration, and heating of 
the electrons. We will follow the evolution of the kinetic energy 
distribution in time and calculate from these distributions 
the hard X-ray (HXR) and microwave spectra,
which can be compared to the observations. 
Moreover, we will
shortly inquire the case of ion heating and acceleration.

%In section 2, we discuss the wider context of the acceleration
%problem and refer to the different approaches which had been made
%in order to solve it. 
The basic elements of our model are presented
in Section 2. In
section 3, we present some of our results, and in section 4 we discuss our
findings and propose ways for continuing the exploration of the
problem along the path of our modeling approach. A conclusion is presented 
in section 5.

\section{The model\label{Sec3}}

\begin{figure}[ht]
  \centering
  \epsfig{file=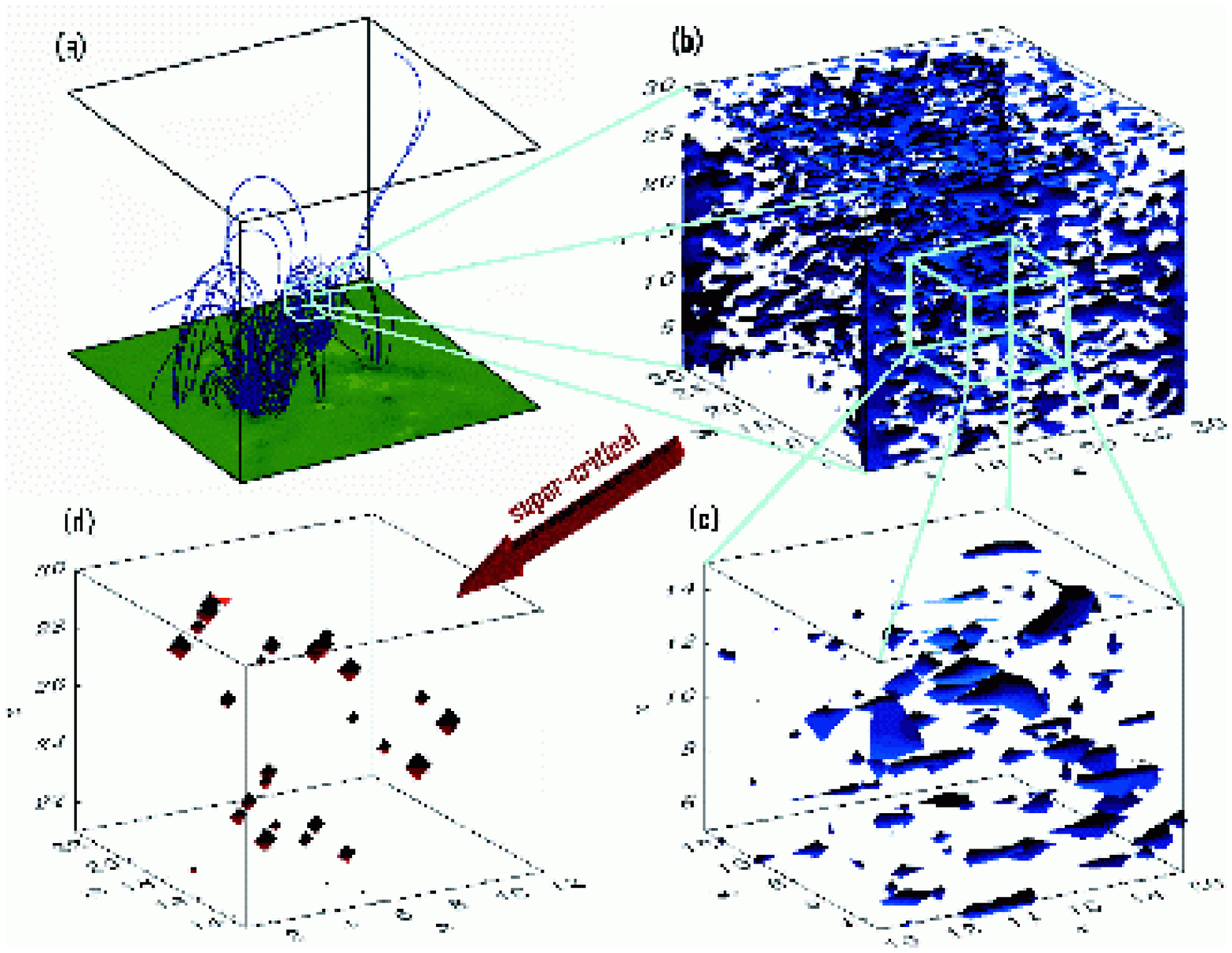,height=10.2cm}
  \caption{(a) Simulated magnetogram of a photospheric active region 
   and force-free magnetic field-lines, extrapolated into the corona
    (generated by the model of \citet{Fragos03}). ---
   (b) Sub-critical current iso-surfaces in space, as yielded by the X-CA 
   model, which models a sub-volume of a coronal active region. ---
   (c) The same as (b), zoomed.
   (d) Temporal snap-shot of the X-CA model during a flare, showing the 
   spatial distribution of the UCS (super-critical current iso-surface) 
   inside the complex active region.
   The UCS form a fractal, and they are connected through large-scale 
   magnetic field lines. 
   The connectivity between the UCS
   is an important ingredient of the acceleration mechanism proposed.
   \label{plotone}}
\end{figure}

In the model we propose in this article, a 3-D large scale magnetic
topology, close to the structures estimated by force-free
extrapolations of the photospheric magnetic fields, is the back bone
of the system (see Fig.\ \ref{plotone}(a)). The magnetic field lines are 
the carriers of the
high energy particles. The UCS, which basically
are magnetic discontinuities, appear sporadically inside
the large scale magnetic topology, and they represent the
dissipation areas of the turbulent system (see Fig.\ \ref{plotone}(d)).
The characteristics of the UCS, such as
size, energy content, etc., are different from UCS to UCS, following
probabilistic laws which in principle should be dictated by the data,
most of them though are not directly observable.
The UCS are assumed to be small-scale regions, distributed over 
space in a complex way. We thus implement the scenario of fragmented 
energy release, based on Parker's original conjecture \citep{Parker88} 
and on the observational evidence of fragmentation in radio spike emission 
\citep[see e.g.][]{Benz03}, and in type III radio-emission 
\citep[e.g.][]{Benz94}.

One way of modeling the appearance, disappearance, and spatial organization 
of UCS inside a large scale topology is with the use of the Extended Cellular
Automaton (X-CA) model \citep{Isl98,Isl00,Isl01}. Fig.\ \ref{plotone} 
illustrates some basic features of the X-CA model.

%\clearpage
%\clearpage

The X-CA model has as a core a cellular automaton model of the
sand-pile type and is run in the state of Self-Organized
Criticality (SOC). It is extended to be fully consistent with MHD:
The primary grid variable is the vector-potential, and the
magnetic field and the current are calculated by means of
interpolation as derivatives of the vector potential in the usual
sense of MHD, guaranteeing $\nabla\cdot \vec B=0$ and $\vec
J=(4\pi/c)\cdot\nabla\times\vec B$ everywhere in the simulated
3-dimensional volume. The electric field is defined as $\vec
E=\eta \vec J$, with $\eta$ the resistivity. %diffusivity. 
The latter usually is
negligibly small, but if a threshold in the current is locally
reached ($\vert\vec J\vert>J_{cr}$), then current driven
instabilities are assumed to occur, $\eta$ becomes anomalous in
turn, and the resistive electric field locally increases
drastically. These localized regions of intense electric fields
are the UCS in the X-CA model.

The X-CA model yields distributions of total energy and peak flux
which are compatible with the observations. The UCS in the X-CA
form a set which is highly fragmented in space and time, the
individual UCS are small scale regions, varying in size, and are
short-lived (see Fig.\ \ref{plotone}(d)). They do not form in their ensemble 
a simple
large scale structure, but they form a fractal set with fractal
dimension roughly $D_F=1.8$ \citep{Isl03b,Mc02}. 
The individual UCS also do usually not split
into smaller UCS, but they trigger new UCS in their
neighborhood, so that different chains of UCS travel through the
active region, triggering new side-chains of UCS on their way.

\begin{figure}[ht]
  \centering
  \epsfig{file=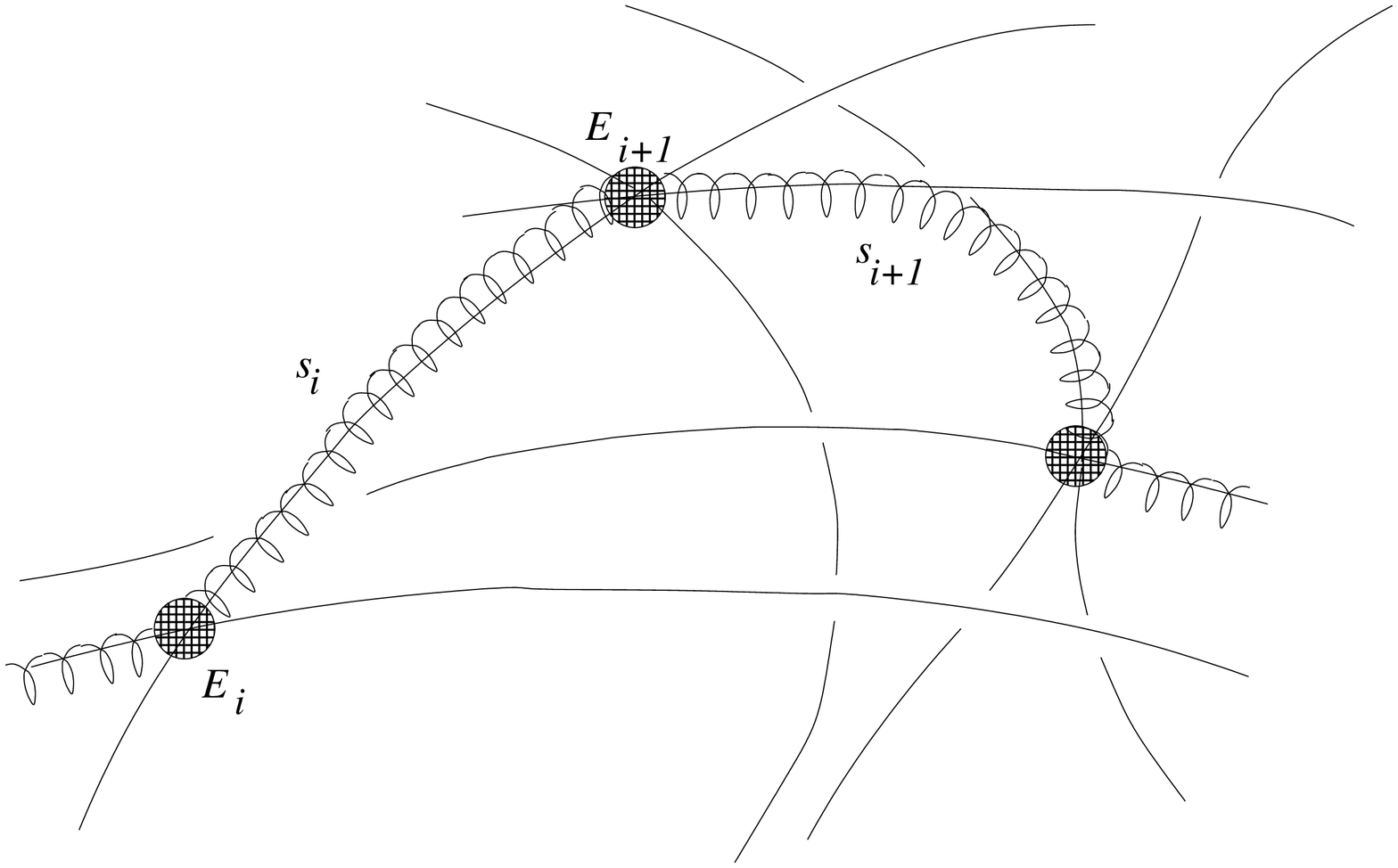,height=6cm}
  \caption{Sketch of the basic elements of the considered model:
   A particle (spiraling line) follows basically the magnetic field
   lines (solid lines), undergoing though also drifts, and 
  travels in this way freely a distance $s_i$, until it enters
  a UCS (filled circles), where it is accelerated by the associated 
  effective DC electric field $E_{i+1}$. After the acceleration event, the 
  particle again moves freely until it meets a new UCS.
   \label{plottwo}}
\end{figure}

Following the picture we have from the X-CA model, we consider in
this article that the UCS act as nodes of activity (localized
accelerators) inside a passive 3-D large scale magnetic topology.
The UCS are short lived and appear randomly inside the large scale
magnetic topology when specific conditions for instability are
met. The sketch in Fig.\ \ref{plottwo} illustrates the situation.

%\clearpage
%\clearpage

Modeling this dynamic accelerator requires the knowledge of three
probability density functions:

%\begin{itemize}
%  \item 

\noindent (1) The probability density $P_1(s)$ defines the
  distances a charged particle travels freely in-between two
  subsequent encounters with a UCS. The series of distances
  $s^{(j)}_1,s^{(j)}_2, ... s^{(j)}_n,...$, generated by the probability
  density $P_1(s)$, characterizes the trajectory of the $j$th particle 
  in space. Every particle follows a different characteristic path.

  The probability density $P_1(s)$ relates the particle acceleration process 
  to the large scale topology by implicitly representing the effects of the 
  topology (the magnetic topology itself does not appear explicitly in the 
  model). This aspect was never taken into account in previous 
  acceleration models.

%  \item 
\noindent  (2) The probability density $P_2(E)$ provides the
  effective electric field $E^{(j)}_i$ acting on the $j$th particle for the
  effective time $\tau^{(j)}_i$ it spends inside the $i$th UCS.
  Particles follow very complicated trajectories inside the UCS. They may be
  accelerated by more than one acceleration mechanisms,
  but what actually is important for our model is the final outcome,
  i.e.\ we characterize a UCS as a simple input-output system, in which
  an effective DC electric field is acting.
  We assume thus that the effective action of a UCS is to increase
  a particle's momentum by
  $\Delta\vec p^{(j)}_i = e\, \vec E^{(j)}_i \, \tau^{(j)}_i$.

%  \item 
\noindent (3) Finally, for the third probability, we have two alternative 
  choices:
  either we give the probability density $P_{3a}(\tau)$ of the
  effective times $\tau^{(j)}_i$ a UCS interacts with a charged particle,
  or we prescribe the probability density $P_{3b}(\ell)$ of the effective 
  acceleration lengths $\ell^{(j)}_i$, i.e.\  the lengths of the trajectories 
  along which the particles are accelerated.

%\end{itemize}
The probabilities $P_1$, $P_2$, and $P_3$ (either $P_{3a}$ or $P_{3b}$) 
are actually dependent on time,
they must be assumed to change in the course of a flare, reflecting the
changes in the overall density and connectivity of the UCS --- 
the flaring active region is an evolving network of UCS,
whose instantaneous state determines the instantaneous intensity of the flare. 
We will though concentrate in our modeling on a short time during a flare,
typically one second, so that we can assume the state of the active region
not to change significantly, and $P_1$, $P_2$, and $P_3$ can be considered
independent of time.  

We will show next that the above probabilities define the
charged particle dynamics inside the flaring region.

\subsection{Charged particle dynamics\label{Sec31}}

The particle $j$ starts with initial momentum $\vec p^{(j)}_0$ from the
initial position $\vec r^{(j)}_0=0$ at time $t=0$. The initial
momentum $\vec p^{(j)}_0$ is such that the corresponding velocity
$\vert \vec v^{(j)}_0 \vert$
is drawn at random from the tail of a Maxwellian,
$\vert \vec v^{(j)}_0\vert \geq v_{th}$,
with $v_{th}$ the thermal velocity. The particle is assumed
to find itself in the neighborhood of a UCS at time $t=0$, enters it
immediately and undergoes a first acceleration process.

During an interaction with a UCS, the particle's momentum in
principle evolves according to
\begin{equation}
\frac{d\vec p^{(j)}}{dt} = e\vec E + \frac{e}{c} \vec v^{(j)} \times \vec B
\label{dpdt0}
\end{equation}
where $\vec p^{(j)} =\gamma m \vec v^{(j)}$ is the particle's relativistic
momentum ($\gamma =1/\sqrt{1-v^2/c^2}$, with $v=\vert \vec v^{(j)} \vert$),
$e$ is the charge of the particle, and $c$ the speed of light.
Inside the UCS, $\vec
E$ and $\vec B$ are complex functions of time along the particle's
complex trajectory. 
We thus average --- in a loose sense ---
%In a loose sense, Eq.\ (\ref{dpdt}) can be considered  as the average of 
Eq.\ (\ref{dpdt0}) over many internal trajectories in a given
UCS, assuming that the corresponding averages exist,
\begin{equation}
\frac{d\langle \vec p\rangle }{dt}
= e\langle \vec E\rangle +
            \frac{e}{c} \langle \vec v \times \vec B \rangle .
\label{h2}
\end{equation}
$\langle \vec E \rangle$ is now constant in a given UCS,
and assuming that $\vec B$ and above all $\vec v$ vary wildly in
direction, so that $\langle \vec v \times \vec B \rangle =0$ is a
reasonable assumption, leads to % Eq.\ (\ref{dpdt}).
\begin{equation}
  \frac{d\langle\vec p\rangle}{dt} = e \langle\vec E\rangle ,
\label{dpdt}
\end{equation}
and we have reduced 
the UCS to a simple constant DC electric field device.
Eq.\ (\ref{dpdt}) is of course a strong simplification of the internal UCS 
physics, it actually implies that we
treat the UCS as black-boxes, concentrating just on the
statistical law of how the output is related to the input.
The electric field $\langle\vec E\rangle$ must consequently be considered
as an {\it effective} electric field, which summarizes the complex
effects of shock-waves, turbulence, and true DC-electric fields
that must be expected to appear in and at a UCS (see Sec.\ 
\ref{Sec1}). 

The effective electric field is constant in a UCS, as mentioned,
it is though assumed to vary from UCS to UCS in a stochastic way, as
prescribed by the probability density $P_2(E)$, where $E$ stands
for the magnitude of the effective electric field $\langle\vec E\rangle$.

After all, during the interaction with the $i$th UCS, a particle's
momentum increases from $\vec p^{(j)}_i$ to $\vec p^{(j)}_{i+1}$ according to
\begin{equation}\label{acc}
  \vec p^{(j)}_{i+1} = \vec p^{(j)}_i + 
              e \langle\vec E^{(j)}_i\rangle\cdot \tau^{(j)}_i ,
\label{pincr}
\end{equation}
where $\langle\vec E^{(j)}_i\rangle$ is generated by the probability density 
$P_2(E)$. To determine $\tau^{(j)}_i$, we have two options in the model:
In the first case, we directly generate $\tau^{(j)}_i$ from the probability 
density $P_{3a}(\tau)$. In the second case, the primary random variable is 
the acceleration length $\ell^{(j)}_i$, which is generated by 
the probability density $P_{3b}(\ell)$, and $\tau^{(j)}_i$ is derived 
under the assumption that the particle performs a relativistic, one-dimensional
motion along the electric field of magnitude $E^{(j)}_i$ and length
$\ell^{(j)}_i$, with initial momentum the magnitude of the total momentum 
$p^{(j)}_i$ the particle has before entering the UCS. 
%These assumptions yield
%\begin{equation}
%\tau^{(j)}_i = 
%\left[ \left\{ \frac{\ell}{c}+\frac{mc}{eE^{(j)}_i} 
%                   \left( 1+\left(\frac{p^{(j)}_i}{mc}\right)^2 \right)^{1/2}
%\right\}^2 - \left(\frac{mc}{eE^{(j)}_i}\right)^2
%\right]^{1/2} - \frac{p^{(j)}_i}{eE^{(j)}_i}
%\label{tau_ell}
%\end{equation}
%for the second case.
In this second case then, the time a particle spends inside
a UCS is reduced when the particle's velocity increases.  

After the particles has left the UCS, it performs a free flight
until it meets again a UCS and undergoes a new acceleration
process (see Fig.\ \ref{plottwo}). The probability density $P_1(s)$ determines 
the spatial distance $s^{(j)}_i$ the particle travels before it meets this next
UCS, situated at
\begin{equation}
\vec r^{(j)}_{i+1} = \vec r^{(j)}_i + s^{(j)}_i\,\hat{r}^{(j)}_{i+1} ,
\end{equation}
where $\hat{r}^{(j)}_{i+1}\equiv \vec p^{(j)}_{i+1} / 
                                  \vert \vec p^{(j)}_{i+1} \vert$
is a unit vector into the direction of the free flight, and
$\vec r^{(j)}_i$ is the location of the previous UCS the particle had met.
Note that the spatial displacement a particle undergoes inside a UCS is
neglected, corresponding to our assumption that the UCS are small scale
regions of negligible extent.

We keep track of the time passed during the acceleration process and the
free flight,
\begin{equation}\label{tim}
  t^{(j)}_{i+1} = t^{(j)}_{i} +\tau^{(j)}_{i} + s^{(j)}_{i}/v^{(j)}_{i+1},
\end{equation}
where $t^{(j)}_{i+1}$ is the time when the particle enters the
$i+1$th UCS.

The particle starts a new cycle of acceleration and free flight at
this point, the process as a whole is a cyclic one with continued
probabilistic jumps in position- and momentum-space.
In this article, we will monitor the system for times which are relatively
short, of the order of one second. For such times, the particles can be
assumed to be trapped inside the
overall acceleration volume $V_{acc}=L^3_{acc}$,
an assumption which will be confirmed by the results we present below
in Sec.\ \ref{Sec4}.
We thus do not have to include in the model the loss of particles
which leave the active volume. 

Let us now define the probability densities $P_1,\, P_2,\, P_3$ used in this
study.

\noindent
\textit{The probability density $P_1(s)$ of jump increments:}
The active, flaring region may be assumed to be in the state of
MHD turbulence, embedded in a complex, large scale magnetic
topology. We claim that the UCS, i.e.\ the regions of dissipation,
are distributed in such a way that they form in their ensemble a
fractal set. This claim is based on two facts: (i) Flaring active
regions have successfully been modeled with Self-Organized
Criticality \citep[SOC; see e.g.][]{Lu91,Lu93,Isl00,Isl01}. It was
demonstrated in \citet{Isl98,Isl00,Isl01} that the unstable sites
in the SOC models represent actually small scale current
dissipation regions, i.e. they can be considered as UCS.
Furthermore, \citet{Mc02} and \citet{Isl03b} have shown that the regions of
dissipation in the SOC models at fixed times form a fractal, with
fractal dimension roughly $D_F=1.8$. (ii) From investigations on
hydrodynamic turbulence we know that the eddies in the inertial
regime have a scale free, power-law size distribution, making it
plausible that at the dissipative scale a fractal set is formed,
and indeed, different experiments let us conclude that the dissipative
regions form a fractal with dimension around $2.8$
\citep[see][and references therein]{Anselmet01}.

The particles in our model are thus assumed to move from UCS to
UCS, the latter being distributed such that they form a fractal
set. \citet{Isl03} analyzed the kind of random walk where
particles move in a volume in which a fractal resides, traveling usually
freely but being scattered (accelerated) when they encounter a
part of the fractal set. They showed that in this case the
distribution of free travel distances $s$ in-between two
subsequent encounters with the fractal is distributed in good
approximation according to
\begin{equation}
p(s)\propto s^{D_F-3}
\end{equation}
as long as $D_F<2$. For $D_F> 2$, $p(r)$ is decaying
exponentially. Given that a dimension $D_F$ below two is reported
for SOC models, as stated above, we are led to assume that
$P_1(s)$ is of power-law form, with index between $-1$ and $-3$,
preferably near a value of $D_F-3=-1.2$ 
\citep[with $D_F=1.8$, according to][]{Isl03b,Mc02}. 
Not included in the study of
\citet{Isl03} are two effects, (a) that the particles do not move
on straight line paths in-between two subsequent interactions with
UCS, but they follow the bent magnetic field lines, and (b) that
particles can be mirrored and trapped in some regions, making in
this way the free travel distances larger. It is thus reasonable
to consider the power-law index of $P_1(s)$ as a free parameter.

After all, we assume that the freely traveled distances $s$ are
distributed according to
\begin{equation}\label{P1}
P_1(s)=A\, s^{-a},\;\;\textrm{with} \;\;l_{min}<s<l_{max},
\end{equation}
where $l_{min}(L_{acc})$ and $l_{max}(L_{acc})$ are related to the
characteristic length of the coronal active region $L_{acc}$,
and $A$ is a normalization constant.

\noindent
\textit{The probability density $P_2(E)$ of effective electric fields:}
The second probability density determines the effective electric
field attached to a specific UCS. Its form should in principle
be induced from a respective study, either from observations,
which is not feasible, so-far, or from the simulation and
modeling of a respective set-up, which to our knowledge seems not
to exist, up to date. We are thus forced to try large classes of
distributions. Two cases of distributions are definitely of
particular interest, the 'well-behaved' case, where $P_2$ is
Gaussian, and the 'ill-behaved' case, where $P_2$ is of power-law
form.
The Gaussian
case is well behaved in the sense that all the moments are finite,
and it is a reasonable choice because of the Central Limit
Theorem, which suggests Gaussian distributions if the electric field is
the result of the superposition of many uncontrollable, small
processes. The power-law case is ill-behaved in the sense that
the moments on from a certain order (depending on the 
power-law index) are infinite. It represents the case of scale-free
processes, as they appear for instance in SOC models. A
characteristics of power-law distributions is the importance of
the tail, which in fact causes the dominating effects.

Trying also the case of Gaussian distributions and guided by the
results, we present in this study only the case where the
distribution of the electric field's magnitude is of power-law form,
\begin{equation}\label{eff}
  P_2(E)=B\,E^{-b},\;\;\; \textrm{with} \;\; E_{min}<E<E_{max}
\end{equation}
which shows better compatibility with the observations. We just
note that most acceleration mechanisms mentioned earlier have
power-law probability distributions for the driving quantity.
$E_{min}(E_D)$ and $E_{max}(E_D)$ are related to the Dreicer field $E_D$,
and $B$ is the normalization constant. 
The Dreicer field is the electric field that leads an electron with 
initial velocity
equal to the thermal velocity to unlimited acceleration (assuming
that only Coulomb collisions can provide the energy losses), and
for typical solar parameters it is $\sim 10^{-2}\,$V/m. Assuming
though an anomalous collision frequency due to the presence of low
frequency waves or of turbulent magnetic fields, $E_D$
increases dramatically by several orders of magnitude. 
We note that choosing the 
Dreicer field as a reference value is somewhat arbitrary, we might as well
have used a typical coronal convective electric field. 
The effective electric field is then determined as
$\langle\vec E\rangle = E\,\hat{r}$, where $\hat{r}$ is a 3-D unit 
vector into a completely random direction.

\noindent
\textit{The probability densities $P_{3a}(\tau)$ and $P_{3b}(\ell)$ of 
effective acceleration-times and -lengths, respectively:}
To complete the description of the acceleration process, we have two 
choices, we either prescribe the acceleration lengths
or the acceleration times. Also for both these distributions a model would be
needed. Since the acceleration-times and -lengths appear though only 
in combination with the electric fields in the momentum increment,
$e \langle\vec E^{(j)}_i\rangle\cdot \tau^{(j)}_i$ [see Eq.\ (\ref{pincr});
the acceleration-lengths come in indirectly], 
%through Eq.\ (\ref{tau_ell})], 
we can absorb any non-standard feature, such a scale-freeness or other strong
non-Gaussianities, in the distribution of $E$. This is also
reasonable since all three, $\tau$, $\ell$ and $E$, are effective quantities.

We assume thus in the first variant of the model that the time a particle 
spends inside a UCS obeys
a Gaussian distribution with mean value $\tau_c$ and standard
deviation $\tau_m$,
\begin{equation}\label{Pt}
P_{3a}(\tau)=C_a e^{-\frac{(\tau-\tau_c)^2}{2 \tau^2_m}} .
\end{equation}
In the second variant, we prescribe the distribution of the 
acceleration-lengths, and the 
acceleration times are calculated as secondary quantities.
%through Eq.\ (\ref{tau_ell}).
Again, a Gaussian distribution is assumed,
\begin{equation}\label{Pl}
P_{3b}(\ell)=C_b e^{-\frac{(\ell-\ell_c)^2}{2 \ell^2_m}} ,
\end{equation}
with mean $\ell_c$ and standard deviation $\ell_m$.
Defined in this way, the acceleration-times or -lengths are
not essential for the acceleration process, they influence though
the overall acceleration time-scale, i.e.\ the global timing of
acceleration.

It should be noted that the effective 
acceleration length, which is the length of the part of the internal 
trajectory along which a particle is accelerated,
is very unlikely to be just equal to the linear size of the UCS, since a 
rather complex internal trajectory must be expected inside and near the UCS,
including phenomena like trapping at the UCS and reinjection. 
Thus, the distribution of acceleration-lengths used in our model does 
not give insight into the size-distribution of the 
UCS. The acceleration times, on the other hand, 
can be compared to observed phenomena (see Sec.\ \ref{Discussion}).
Yet, prescribing the acceleration lengths causes that 
the faster a particle is, the less time it spends in the acceleration 
events, which is physically reasonable.

\medskip

Eqs.\ (\ref{acc})-(\ref{Pl}) allow to follow the evolution of
charged particles inside an evolving network of UCS. Our main
interest in the simulations which we will present is to follow the
evolution of the distribution in kinetic energy ($E_{kin} = (\gamma-1)mc^2$)
as a function of time. Thereto, we monitor the kinetic energies of the
particles at prefixed times, and construct their distribution
functions $p(E_{kin},t)$ (normalized to 1).

%\subsection{HXR and microwave emission}

%\subsubsection{HXR Bremsstrahlung\label{Sec321}}

\noindent\textbf{HXR Bremsstrahlung:}
We assume that the distribution $p(E_{kin},t)$ of particles, as it is
formed at time $t$, precipitates onto lower layers of the solar
atmosphere, where it emits Bremsstrahlung while being completely
thermalized. As usual, we assume thick-target Bremsstrahlung.
The HXR emission spectra are calculated by using the formalism of
\citet{Brown1971}, according to which the photon count rate
$I(\epsilon)$ (photons of energy $\epsilon$ per unit time and unit energy range)
is calculated through an integral 
%\begin{eqnarray}
%I(\epsilon) &=& A C\,\frac{1}{\epsilon}
%  \int\limits_\epsilon^\infty \!\! dE_{kin}\,F(E_{kin})  \nonumber \\
% && \times
%  \int\limits_\epsilon^{E_{kin}} \!\!\!  dE\,
%     \log\left(\frac{1+\sqrt{1-\epsilon/E}}{1-\sqrt{1-\epsilon/E}}\right) ,
%\label{brems}
%\end{eqnarray}
%with $C$ a constant \citep[for details see][]{Brown1971}, 
%$\epsilon$ the photon energy, . 
over the flux $F(E_{kin},t)$ of
precipitating electrons, which we determine as
\begin{equation}
F(E_{kin},t) = n_1\, p(E_{kin},t)\,v + n_0\, p_{th}(E_{kin})\,v ,
\end{equation}
with $v$ the velocity corresponding to $E_{kin}$,
$p(E_{kin},t)$ given from the simulations in numerical
form, and $p_{th}(E_{kin})$ is the (normalized to 1) thermal distribution,
taken in analytical form. $n_1$ is the number density
of the precipitating electrons, and $n_0$ the respective density of the
background plasma. To complete the necessary parameters, an emitting area
$A$ has also to be specified.

The integration is done numerically. Since
$p(E_{kin},t)$ is given numerically at a relatively low number
of discrete points --- we use typically $30$ bins in making the
histograms ---, we use the values of $p(E_{kin},t)$ at the midpoints
of the bins and interpolate with cubic splines, which allows for
precise enough numerical integration. The interpolation is
done in the log-log, since we let the bin-size increase with energy
in order to reduce the statistical errors at high energies.
Interpolating in the
linear would lead to strong oscillations at high energies, whereas
interpolation in the log-log is well behaved and follows smoothly the
numerically given data-points.

It is to note that since the particle energy distributions depend
on time, we also get time-dependent HXR spectra, $I(\epsilon,t)$.

%\subsubsection{Gyro-synchrotron microwave emission\label{Sec322}}

\noindent\textbf{Gyro-synchrotron microwave emission:}
During their free flights, the particles gyrate in the 
background magnetic field
and emit synchrotron radiation. Assuming a constant magnetic field 
$B_0$ and a homogeneous source region,
we determine the gyro-synchrotron radiation spectra 
%\begin{equation}
%F_{o,x}(\nu,\theta_0) = \frac{A}{2.238\,10^7} 
%\frac{j_{o,x}(\nu,\theta_0)}{\kappa_{o,x}(\nu,\theta_0)}
%\left\{1-\exp\left[-\kappa_{o,x}(\nu,\theta_0)\,d\right] \right\} ,
%\end{equation}
$F_o(\nu)$ and $F_x(\nu)$, which are the fluxes (in SFU) of the o- and the 
x-mode, respectively.
%$j_o$ and $j_x$ are the respective emissivities, 
%$\kappa_o$ and $\kappa_x$ the respective absorption coefficients,
The respective formulae, including the expressions for the emissivities 
and the absorption coefficients,
are taken from \citet{BenHol92}, and are ultimately based on the work of 
\citet{Ramaty69}. 
The evaluation of these formulae
%$j_{o,x}$ and $\kappa_{o,x}$
involve integrals 
over the distribution $p(\gamma)$ of the relativistic $\gamma$ 
and its first derivative. Our model yields the distribution of kinetic
energies $p(E_{kin})$ in numerical form, which we first transform into the 
distribution
of $\gamma$ according to $p(\gamma)=p(E_{kin})\,dE_{kin}/d\gamma$. 
The resulting $p(\gamma)$ is then given in numerical form at discrete
$\gamma$ values. To integrate over $p(\gamma)$ and to calculate its
derivative, we interpolate $p(\gamma)$ with cubic splines in the log-log,
exactly as we did with $p(E_{kin},t)$ in the case of the HXR emission 
described above. 
The differentiation is done by directly differentiating the interpolating
spline polynomials. 

To completely determine the emission, we have to specify   
the emitting area $A$, the thickness of the source $d$,
and finally the angle $\theta_0$ between the magnetic field and
the line of sight.
The frequency range for which the fluxes are calculated is chosen 
in the microwaves such that it corresponds to the typical observational 
range of current instruments.

\section{Results\label{Sec4}}

The simulations are performed by using $10^6$ particles (electrons, 
and in Sec.\ \ref{ions} protons),
and the system is monitored for $1\,$sec, with the aim to
concentrate our analysis on a short time-interval during the
impulsive phase (for longer times we would have to include
the explicit loss of particles from the accelerating volume).
We performed an extended parametric study, of which we present
here only one particular case, some general comments on the parameter
dependence of the model will be made in Sec.\ \ref{Discussion}.
The applied parameters for the case presented here are:
\begin{itemize}
\item {\it Background plasma parameters:} We assume a
temperature $T\approx 1.2\,10^6\,$K (corresponding to $100\,$eV)
and a density $n_0=10^{10}\,$cm$^{-3}$ for the ambient, background plasma.
To estimate the synchrotron losses, a background magnetic
field $B_0=100\,$G is assumed. For the calculation
of the HXR emission, we use again the afore-mentioned values of the
background plasma temperature $T$ and density $n_0$,
and we moreover assume an emitting area $A=10^{18}\,$cm$^2$
and a density of accelerated particles $n_1=2\,10^7\,$cm$^{-3}$.
The same parameters $B_0$, $n_0$, $n_1$, and $A$ are applied in the 
calculation of the gyro-synchrotron emission, together with a thickness
of the emitting region $d=10^9\,$cm and an angle $\theta_0=45^o$ between
the magnetic field and the line of sight.

\item {\it The Random walk in position space}
is characterized by the power-law exponent $a=1.2$
and by the minimum and maximum jump lengths
$l_{min} = 10^{-6}\,L_{act}$ and $l_{max} = 10^3\, L_{act}$,
with $L_{act}=10^{10}\,$cm [see Eq.\ (\ref{P1})].

\item {\it Electric field statistics:}
The power-law index of the electric field distribution is set to $b=4.5$,
and we choose the range from $E_{min} = E_D$ to $E_{max} = 10^8\,E_D$,
where the Dreicer field $E_D$ is a function $T$ and $n_0$
[see Eq.\ (\ref{eff})].

\item {\it Effective acceleration-times and -lengths:}
Either the acceleration-times or the acceleration-lengths are prescribed. 
The Gaussian of the acceleration-times is determined by the
mean value $\tau_c = 2\,10^{-3}\,$s and the standard deviation
$\tau_m = \frac{2}{3}\,10^{-3}\,$s [see Eq.\ (\ref{Pt})], whereas the 
Gaussian of the acceleration-lengths has mean value $\ell_c=5\,10^7\,$cm 
and standard deviation $\ell_m=\frac{5}{3}\,10^7\,$cm 
[see Eq.\ (\ref{Pl}); we note that the acceleration-lengths cannot be directly
compared to the UCS sizes, see Sec.\ \ref{Sec31}].
\end{itemize}

\subsection{Electron diffusion inside the network of UCS\label{Sec41}}

%\clearpage
\begin{figure}[ht]
 \centering
\epsfig{file=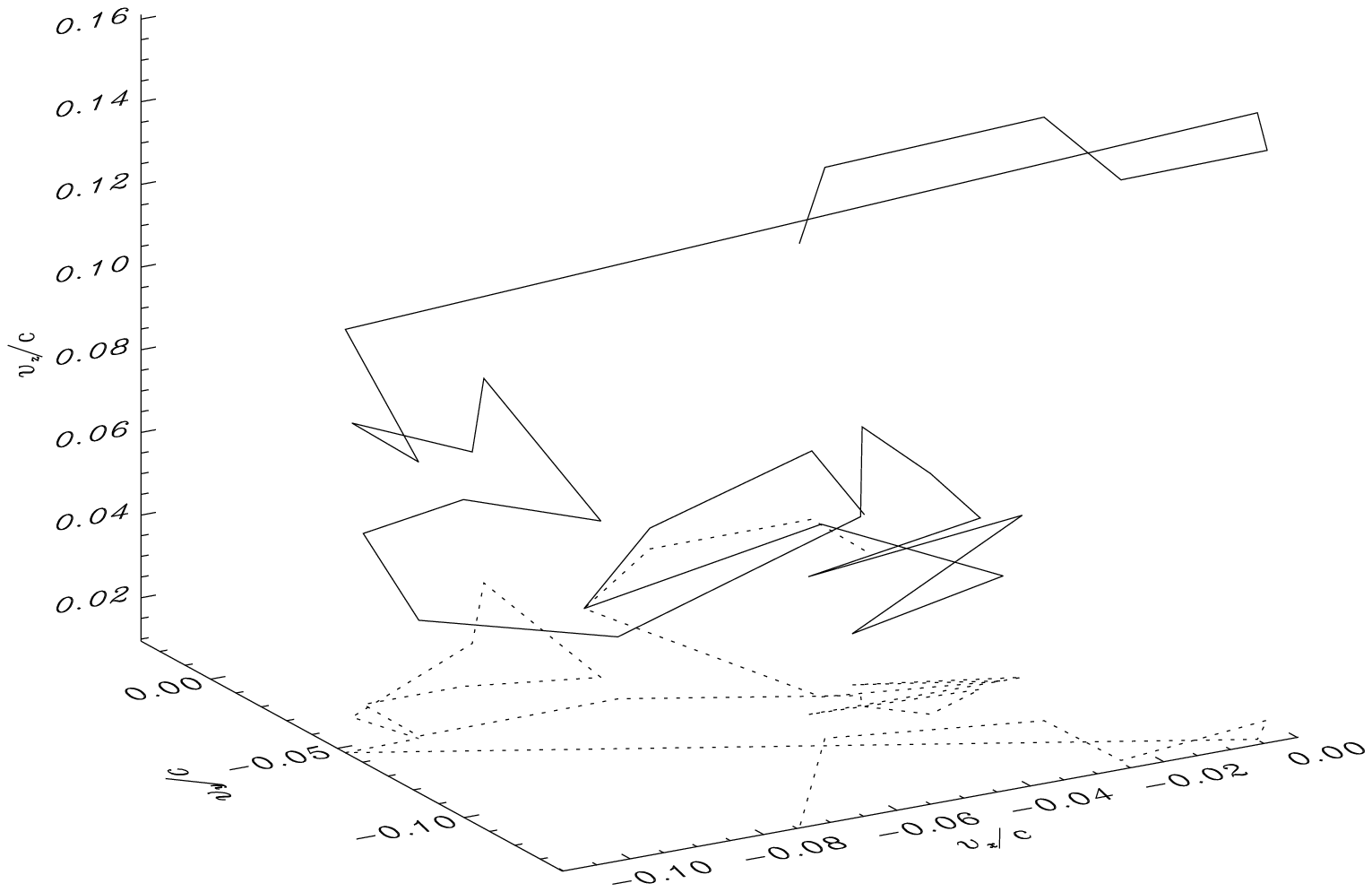,height=6cm}
 \caption{A particle trajectory in $\vec v$-space (solid), together
with its projection onto the bottom ($v_x$-$v_y$) plane (dotted).
The velocities are in units of the speed of light $c$.
\label{plotthree} }
\end{figure}

%\clearpage
\begin{figure}[ht]
 \centering
\epsfig{file=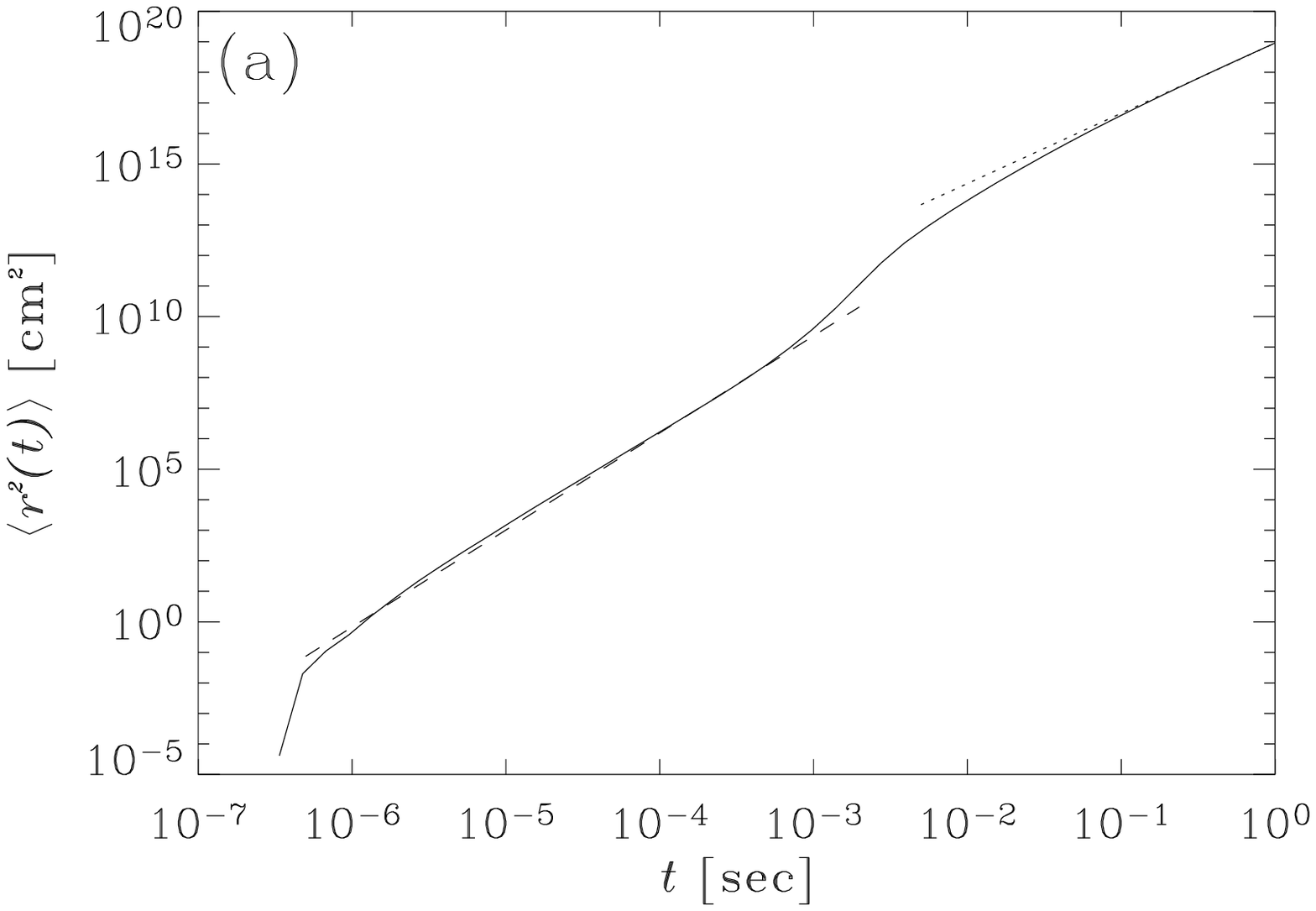,height=6cm}
\epsfig{file=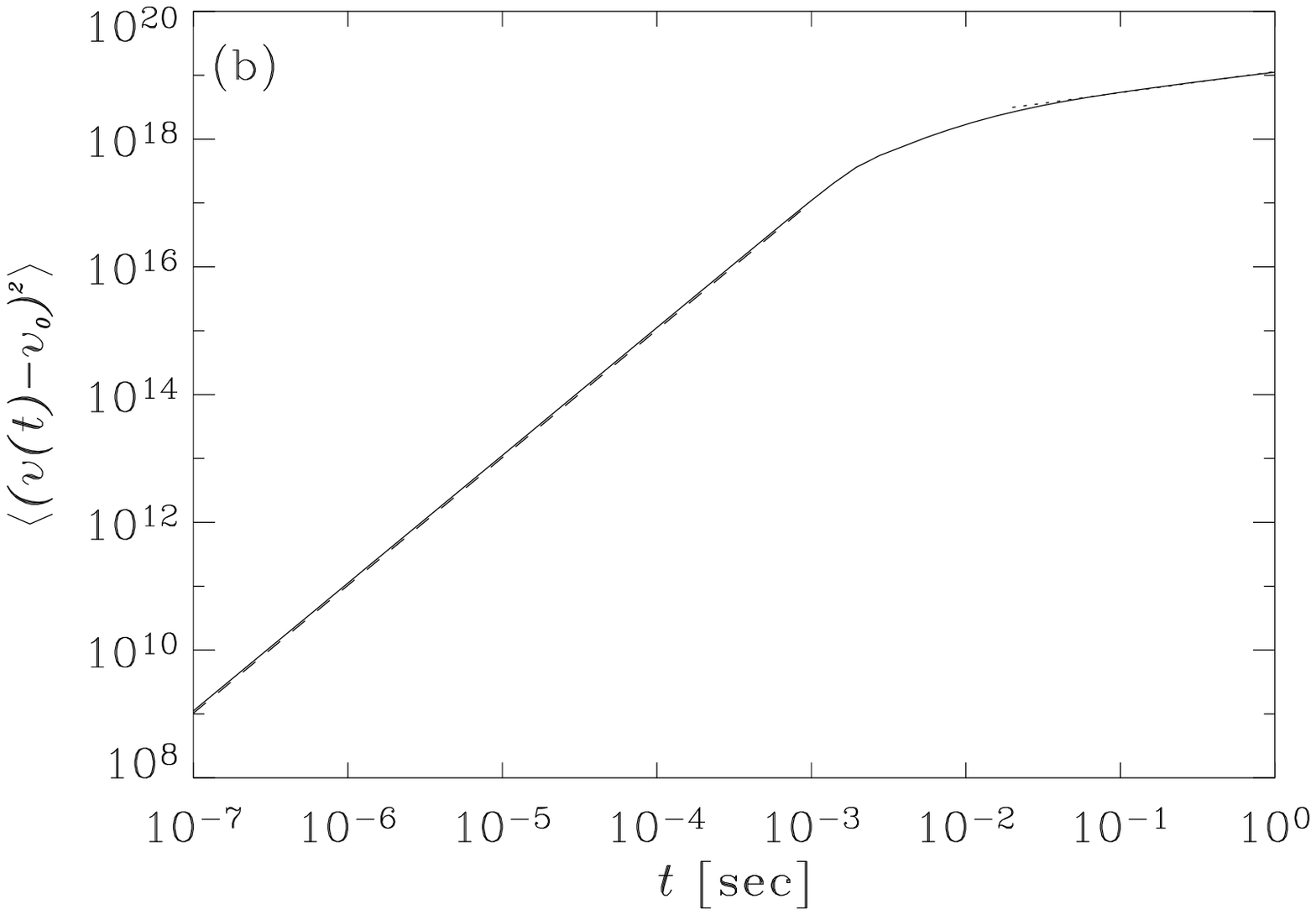,height=6cm}
 \caption{
The diffusive behavior for the case in which the acceleration times
are prescribed:
(a) $\langle r^2(t) \rangle$ vs $t$, together with two reference lines
of slope $3.18$ (dashed) and $2.30$ (dotted), respectively. --- 
(b) $\left\langle (v(t)-v_0)^2 \right\rangle$ vs $t$, with again two 
reference lines of slope $2.00$ (dashed) and $0.33$ (dotted), respectively.
\label{plotfour} }
\end{figure}
%\clearpage

Up to the one second we monitor the system, the electrons
undergo repeatedly acceleration events, whose number is different
from particle to particle: for the variant of the model with the 
acceleration-times prescribed, the minimum number
of acceleration events per particle is found to be $1$,
the maximum is $175$, and the mean is $13.4$.
We thus have quite a low mean number of acceleration events,
and a fraction of the particles just undergoes one, initial, acceleration
process. Fig.\ \ref{plotthree} shows a particle trajectory in 
velocity space. Characteristic is the appearance of jump-lengths 
of various, from small to large, sizes, which is a consequence of the 
power-law form of the momentum increment (see Eqs.\ (\ref{acc}) and 
(\ref{eff})),
and it is actually a typical property of Levy-walks. We do not show
a position-space trajectory, since it is difficult to visualize:
the increments are again power-law distributed (Eq.\ (\ref{P1})), with 
very low index ($1.2$) though, implying that there occur very large 
increments, and in 
a plot just the largest increment will be seen as a straight line,
the smaller increments are not resolved.

To analyze the diffusive behavior of the electrons in position space,
we determine the mean square displacement $\langle r^2(t) \rangle$ of the
particles from the origin as a function of time, and the result is presented
in Fig.\ \ref{plotfour} for the case where the acceleration-times are 
prescribed. For all times the system is monitored, we find strong
super-diffusion, $\langle r^2(t) \rangle \propto t^\gamma$, with $\gamma$
between $2.3$ and $3.2$. The behavior is different
above and below $4\,10^{-3}\,$sec, a time which is related
to the acceleration time: at $\tau_c+3\tau_m=4\,10^{-3}\,$sec
the vast majority of the particles has finished its first acceleration
process (since the acceleration times are Gaussian distributed
[see Eq.\ (\ref{Pt})], 99\% of the particles have an acceleration time
smaller than the time of three standard deviations above the
mean value). Below $4\,10^{-3}\,$sec, the particles typically are still
in their first acceleration process, whereas above $4\,10^{-3}\,$sec,
some particles are on free flights and others are in new
acceleration processes.
We cannot claim that the diffusive behavior has settled
to a stationary behavior in the $1\,$sec we monitor the system.
At $1\,$sec, we find
$\langle r^2(t=1\,sec) \rangle \approx 10^{19}=L_{act}^2/10$, 
the particles have
diffused a distance less than the active region size, so that we do not have
to worry about loosing particles by drifting out of the active region,
as claimed earlier.

We analyze next the diffusive behavior in velocity space, determining
the mean square displacement $\left\langle (v(t)-v_0)^2 \right\rangle$ 
of the particles
from their initial velocity $v_0$. Fig.\ \ref{plotfour} shows the result.
The particles start with $\left\langle (v(t)-v_0)^2 \right\rangle
\propto t^{2.00}$ until roughly 
$0.001\,$sec, a time slightly earlier than $\tau_m$, the mean acceleration
time. This reflects the fact that basically all particles are still
in their first acceleration event, and their velocity evolves according to 
$v(t) \propto (eE/m)t^2$. For larger times, 
$\langle (v(t)-v_0)^2 \rangle$ starts to turn-over, and
in the range from $0.02\,$s to $1\,$s, where all the particles have finished 
their first acceleration event, we 
find $\langle (v(t)-v_0)^2 \rangle \propto t^{0.33}$. The system exhibits thus
clear sub-diffusive behavior in velocity space
(corresponding to the non-relativistic energy space).

\subsection{Kinetic energy distribution, microwave and HXR emission
\label{Sec42}}

%\clearpage
\begin{figure}[ht]
 \centering
\epsfig{file=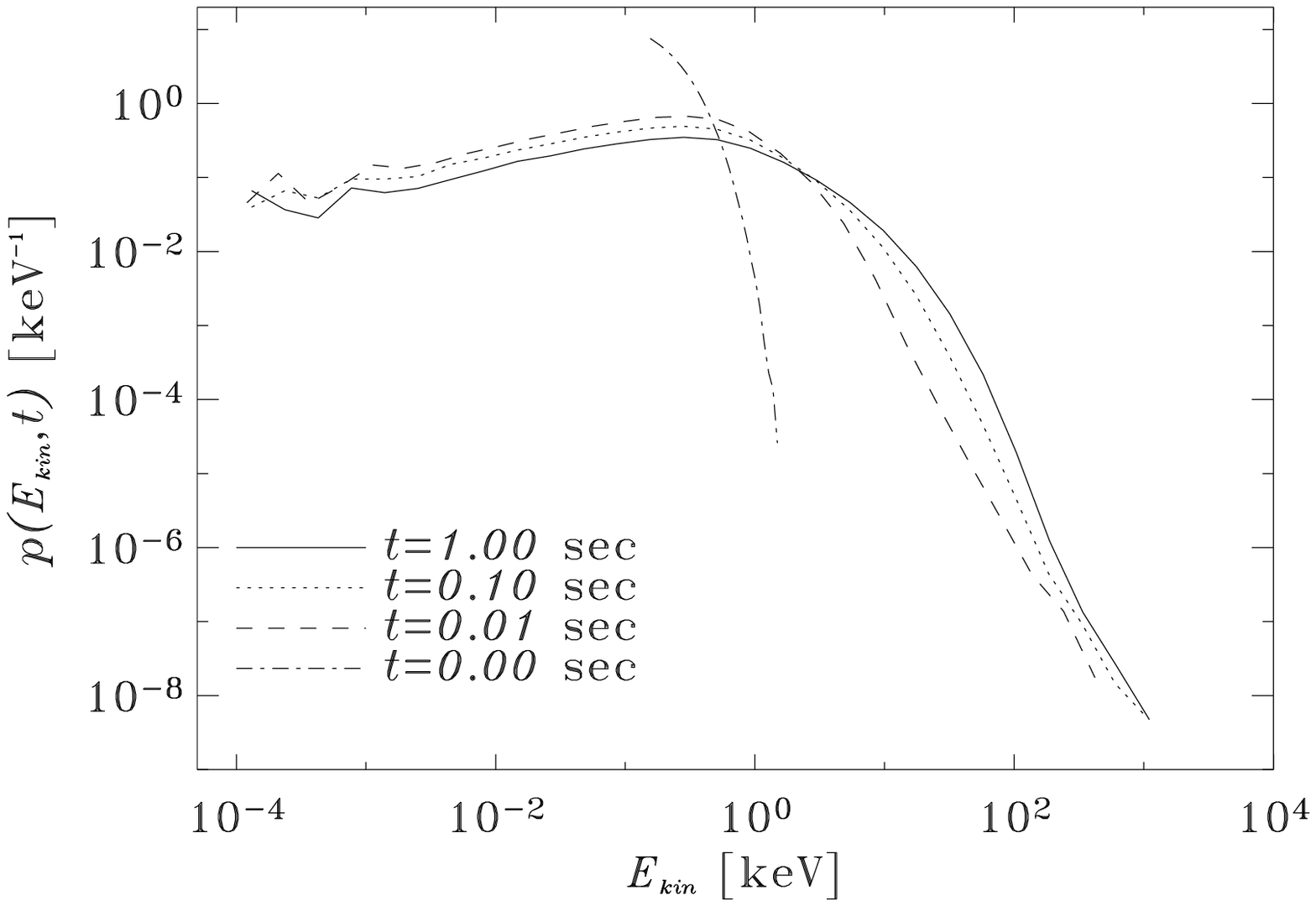,height=6cm}
 \caption{
Kinetic energy distributions $p(E_{kin},t)$ (probability density function,
normalized to one) at times
$t = 0,\, 0.01,\, 0.1,\, 1\,$sec, for the case where the acceleration-times
are prescribed. 
%Top: the acceleration-times are prescribed;
%bottom: the acceleration-lengths are prescribed.
\label{plotfive} }
\end{figure}
%\clearpage

At prefixed times, we collect the kinetic energies
of the electrons and construct their histograms, which, normalized
to $1$, yield the kinetic energy distributions $p(E_{kin},t)$,
shown in Fig.\ \ref{plotfive} for the case with prescribed 
acceleration-times. The distributions remain of similar
shape for the time-period monitored, exhibiting a flatter part at low
energies, and a power-law tail above roughly $5\,$keV. The low energy
part is actually of a Maxwellian type.
The power-law index of the high energy tail varies around $4$,
increasing slightly with time, and the particles also reach higher
energies with increasing time. We see thus a systematic shift of the Maxwellian
towards higher energies, with in parallel the development of a power-law tail
extending to higher and higher energies and steepening.
At $1\,$sec, the most energetic particles have reached kinetic
energies slightly above $1\,$MeV. If we define the approximate 
power-law tail somewhat arbitrarily
to start at $5\,$keV, we find that roughly 35\% of the electrons are
in the tail at $t=1\,$sec. 
The results for the case where we prescribe the acceleration lengths are 
qualitatively similar, the power-law tail is though 
steeper, with a power-law index around $5$.

%\clearpage
\begin{figure}[ht]
 \centering
\epsfig{file=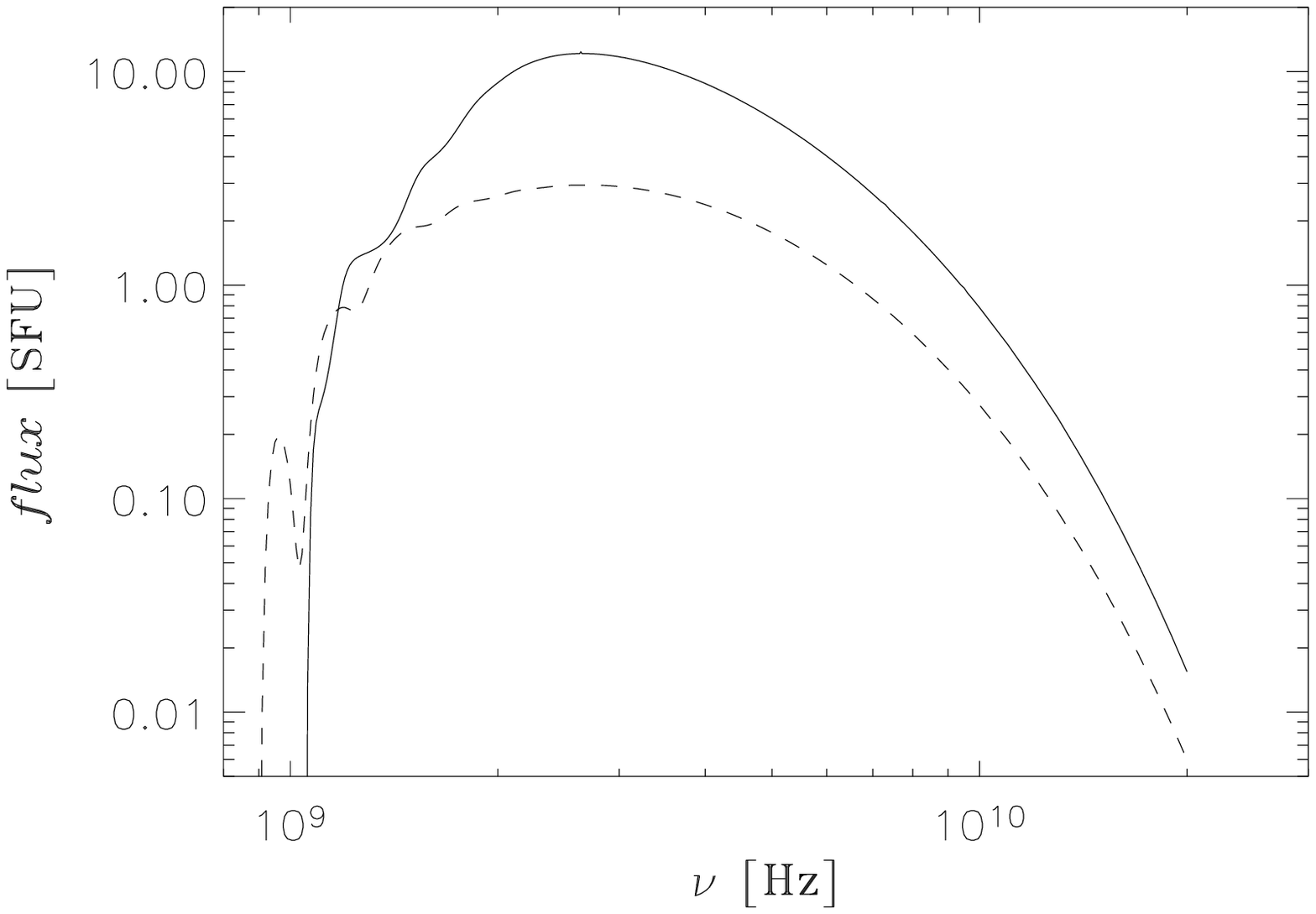,height=6cm}
 \caption{
Microwave-emission spectrum (flux in SFU) at time
$t=1\,$sec; solid: x-mode ($F_x(\nu)$), dashed: o-mode ($F_o(\nu)$).
%Bottom: Degree of polarization, $(F_o-F_x)/(F_x+F_o)$, as a function
%of frequency $\nu$.
\label{plotsix} }
\end{figure}
%\clearpage

During their free flights, the particles move
in a constant background magnetic field, they thus gyrate and emit
synchrotron radiation. The microwave spectrum both for the
x-mode ($F_x(\nu)$) and the o-mode ($F_o(\nu)$) is calculated
according to Sec.\ \ref{Sec31}, and Fig.\ \ref{plotsix}
shows the spectra for the case where the acceleration-times are prescribed. 
Besides some fluctuations at low frequencies,
the spectra exhibit an increase with frequency,
a turn-over, and a decrease towards high frequencies.
The turn-overs for the x- and the o-mode are slightly different,
as are the low-frequency cut-offs. The degree of polarization can 
reach values up to 60\%.

%\clearpage
\begin{figure}[ht]
 \centering
\epsfig{file=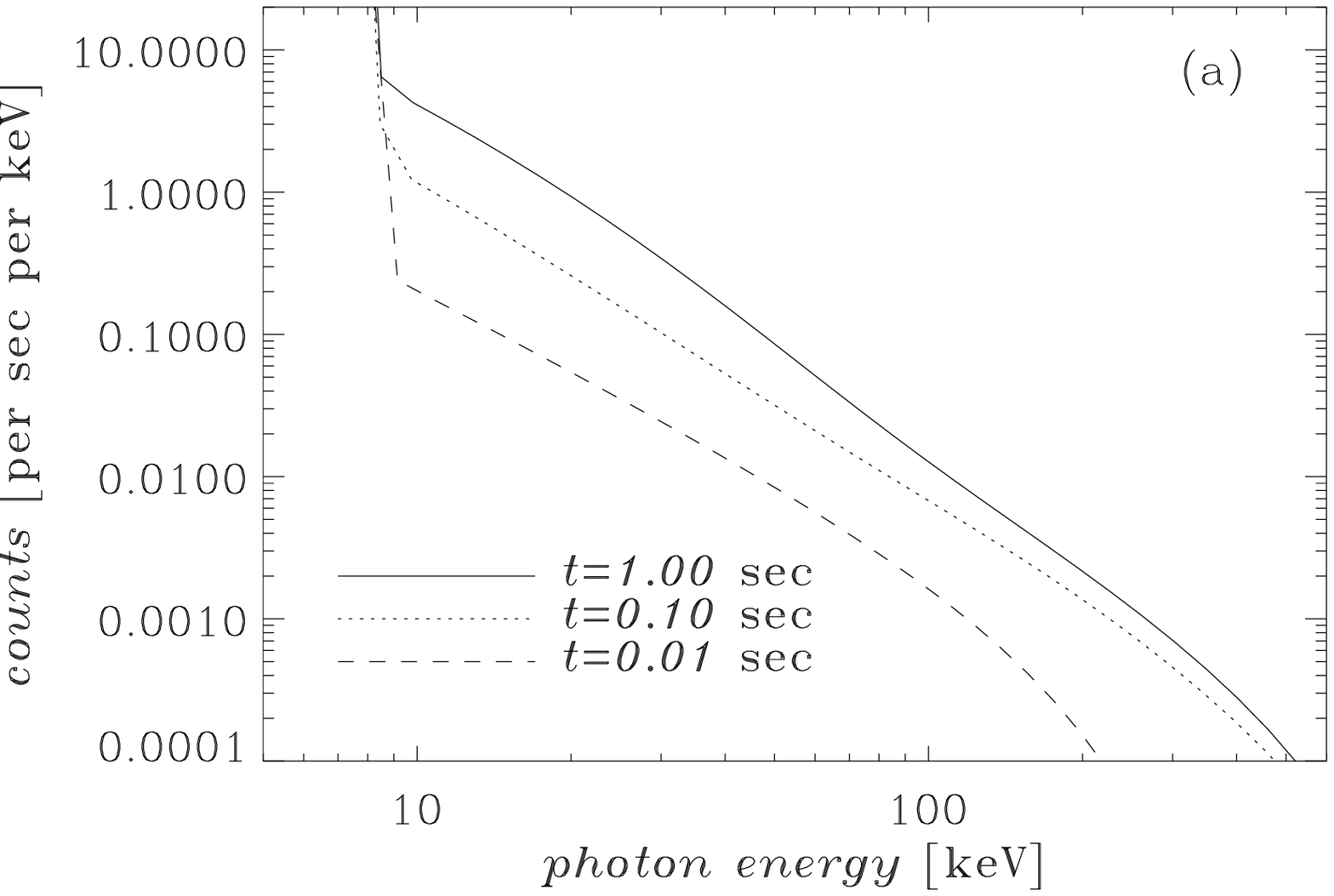,height=6cm}
\epsfig{file=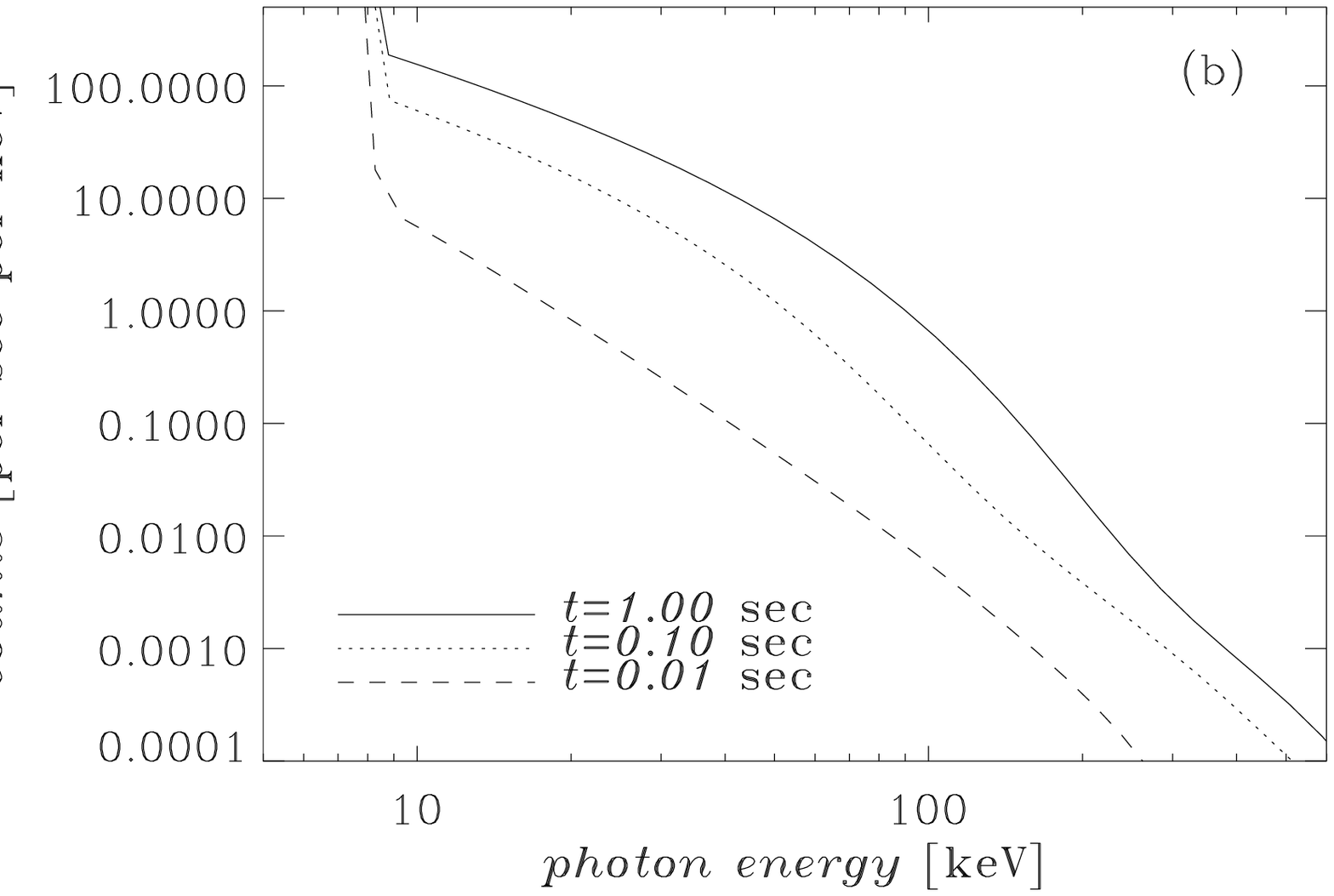,height=6cm}
 \caption{
(a) HXR Bremsstrahlung spectra for the times
$t=0.01,\, 0.1,\, 1\,$sec and with the acceleration-times prescribed,
derived from the corresponding
kinetic energy distributions shown in Fig.\ \ref{plotfive}, 
including thermal emission.
%--- (b) The same as (a), but without thermal emission (and a wider
%energy range plotted).
%In (a) and (b), the acceleration-times had been prescribed. 
--- (b) HXR spectra for prescribed acceleration-lengths.
\label{plotseven} }
\end{figure}
%\clearpage

Next, we let the electron populations, as they are formed at different,
prescribed times, precipitate onto lower layers of the solar atmosphere.
Fig.\ \ref{plotseven}(a) shows the Bremsstrahlung emission spectra as a
function of time, calculated according to Sec.\ \ref{Sec31}, and with 
the acceleration-times prescribed. 
It is to note that Fig.\ \ref{plotseven} shows only the electron contribution,
not included are in particular the ion contribution, the nuclear lines, and 
the emission from particle annihilation.
The lower cut-off in the photon count-rate is chosen to correspond 
to the one of modern instruments, which typically observe power-law decays 
over four orders of magnitude in count rate before reaching the 
noise level \citep[e.g.][]{Lin03}. 
Below the cut-off, the simulated count-rate falls fast off to zero.

For all times monitored, the spectra are of power-law
shape, with a turn-over and fall-off to zero at the photon energy
corresponding to the highest electron energy reached. The steepening
below roughly $15\,$keV is the thermal emission of the background plasma.
The spectral indices 
%are represented graphically in Fig.\ \ref{plotseven}
%(bottom). They 
are all above $2$, and they increase with time,
with a value of $2.1$ at $t=0.01\,$sec and
reaching finally $2.5\,$ at $1\,$sec (from a power-law fit above $10\,$keV).

%Fig.\ \ref{plotseven}(b) shows the HXR emission without the thermal component.
%Above $10\,$keV, the spectra are identical to the ones of of Fig.\ 
%\ref{plotseven}(a). In the range from $1\,$keV to $10\,$keV, the spectra are
%still of power-law shape, the spectral index is now much smaller, though,
%a power-law fit yields a value of $1.5$ at $t=1\,$sec. The thermal 
%emission thus hides the turn-over to flatter non-thermal spectra at lower
%photon energies.

In the case where the acceleration-lengths are prescribed, 
we find for $t=0.01\,$sec a single power-law with index $3.0$, at $t=0.1\,$sec
an approximate double power-law is formed with spectral indices 
$1.9$ at low and $4.3$ at high energies, and finally for $t=1\,$sec
the double power-law persists, with indices $1.7$ at low energies and 
$4.7$ at high energies (Fig.\ \ref{plotseven}(b)). In the course of time,
the power-laws thus get flatter at low energies and steeper at high energies.

We additionally checked the importance of synchrotron radiation losses
for the particle dynamics during the free flights, assuming gyration in a
constant, uniform magnetic field $B_0=100\,$G.
We found that the radiative losses do very insignificantly influence
the results, they basically remain unaltered. The reason is that the
magnitude of the magnetic field is not very large, and above all
the free flight times are too short to allow substantial losses
in energy.

%\noindent
%\textbf{Plasma heating:}
The kinetic energy distributions the model yields are approximately 
Maxwellians with a 
power-law tail above roughly $1\,$keV (Fig.\ \ref{plotfive}). Fitting a 
Maxwellian in the low energy range from $0.01\,$keV
up to roughly $0.5\,$keV yields the temperature of the thermal part
of the distribution.
We find for prescribed acceleration-times that the population of 
electrons moving through the UCS
is heated 
on very fast time scales ($0.01\,$sec) from the initial temperature of 
approximately $1.2\,10^6\,$K to temperatures of the order of $6.6\,10^6\,$K, 
around which it fluctuates until $t=1\,$sec, without any significant further
increase. We have though to note that the Maxwellian does 
not fit perfectly, at low energies there is a systematic over-population. 
This is most likely so since we do not include
collisions, which would thermalize the distribution to an exact Maxwellian.
Nevertheless, the temperatures inferred can be considered as indicative of
the heating process.

\subsection{Acceleration and heating of ions\label{ions}}

%\clearpage
\begin{figure}[ht]
 \centering
\epsfig{file=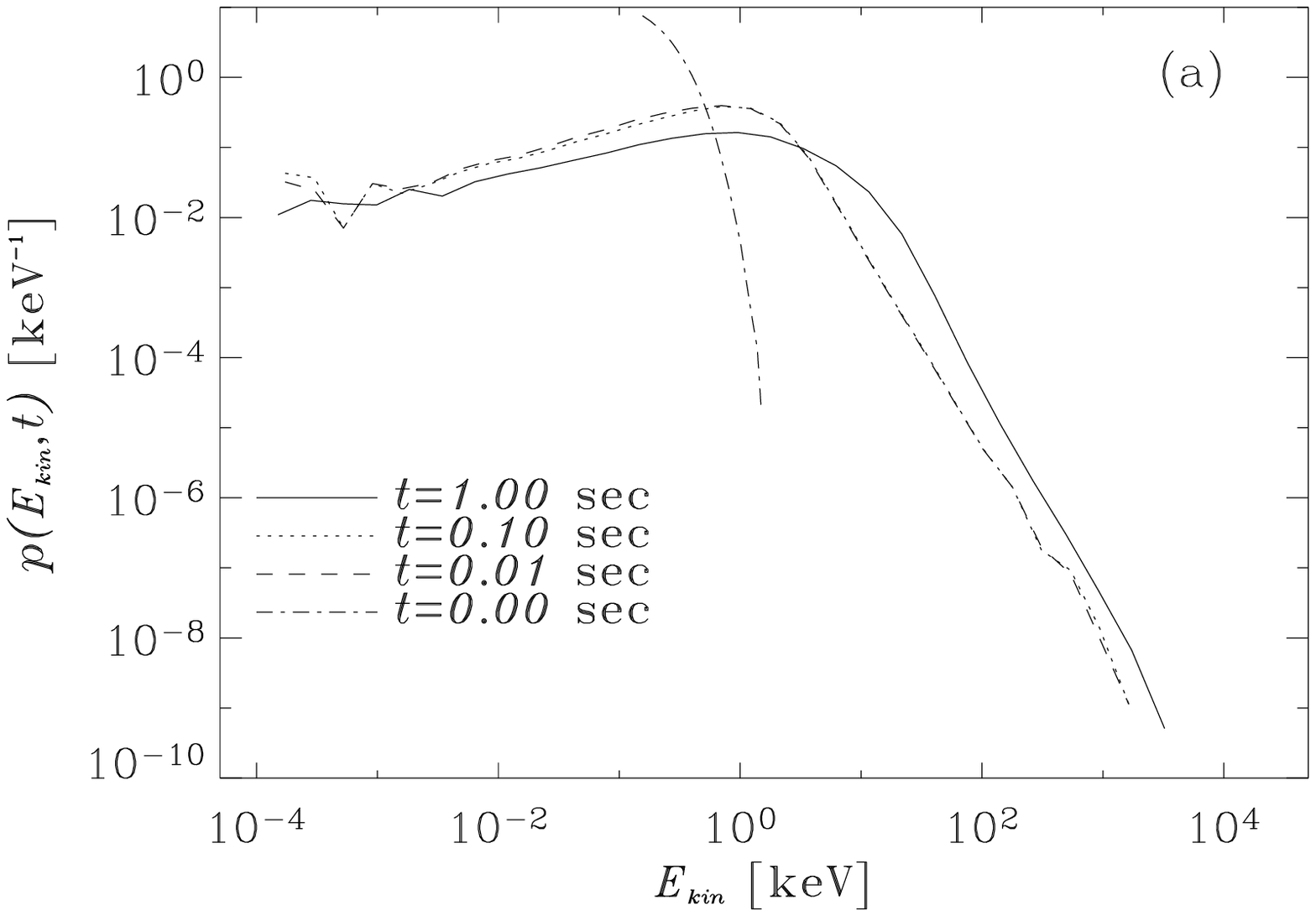,height=6cm}
 \caption{
Kinetic energy distributions (normalized to one) of protons 
at times $t=0,\,0.01,\, 0.1,\, 1\,$sec.
%. --- (a): Applying
%larger acceleration times; --- (b) Applying larger DC electric fields. 
\label{ploteight} }
\end{figure}
%\clearpage

It is of interest to know what will happen to ions which go
through the same kind of process as the electrons. To adjust our
model to the case of ions, we just have to replace the electron mass
with the ion mass. Keeping all the parameters fixed as described in the
beginning of Sec.\ \ref{Sec4} and using the variant of the model with 
prescribed acceleration-times, we find in the case of protons that the
initial distribution is basically unaltered, even for times like
$1000\,$sec. The reason is that the momentum increments are too
small for the ions to undergo a visible change in energy distribution,
they need larger momentum increments. 
There are several possibilities in the frame of our model to achieve 
larger momentum increments, the most straightforward ones are 
either to increase the acceleration times, or 
to increase the effective DC electric field. 

In the case we show here, we increase the acceleration times by a factor of 
$100$ compared to
the electrons, leaving all the other parameters unchanged. We apply thus
the mean value $\tau_c=2\,10^{-1}\,$s and the standard deviation 
$\tau_m=\frac{2}{3} 10^{-1}\,$s in Eq.\ (\ref{Pt}). Fig.\ \ref{ploteight} 
shows the kinetic energy distributions. They exhibit a rough Maxwellian 
at low energies and a 
power-law tail with a slope increasing in time, reaching a value of roughly
$3.2$ at $t=1\,$sec.
The distributions for $t=0.01\,$sec and $t=0.1\,$sec are similar,
most likely since for these times most particles are still in their
first acceleration process, which will have ended for almost 
all particles at the time $\tau_m+3\tau_c=0.4\,$sec (see the corresponding 
argument in Sec.\ \ref{Sec41}). The fraction of particles in the 
tail above $10\,$keV is $29\%$ at $t=1\,$sec.
The Maxwellian part of the distributions
is shifted to higher energies in the course of time, which corresponds to
heating. The model thus yields also heating and acceleration
of the protons. 

We found similar results on adjusting the minimum
of the electric field distribution to $E_{min}=100\,E_D$ (see Eq.\ 
(\ref{eff})),
which increases the mean value of the effective electric field
and causes in this way larger momentum increments for the protons.
%All the other parameters are kept as in case of the electrons.
%The result is shown in Fig.\ \ref{ploteight}(b), the energy distributions
%are again Maxwellians with approximate power-law tails. The index of the
%approximate power-law tail is around $3.5$, and assuming the tail to 
%start at $10\,$keV, we find a fraction of 42\% of the ions to be in the tail
%at $t=1\,$sec.
In the variant of the model with prescribed acceleration-lengths, 
the ions are accelerated up to roughly $550\,$keV, without changing any 
parameter of the model.

\section{Discussion\label{Discussion}}

The model we introduced is the general frame-work of any model of multiple
particle acceleration, it is the most general approach
in the sense that all kinds of multiple acceleration problems,
not just the solar flare problem,
can be cast into the general form of it
by adjusting appropriately the elements.

The model is specified by setting the elements of the spatial
random walk and the random walk in momentum space.
Setting these elements should be done along the guide-line
of physical insight into the problem under study,
an insight which comes from independent studies in different
fields. The nature of the model
partly implies that new kinds of questions
should be asked in the different involved fields, stressing the statistical
nature of different mechanisms and effects.

In the application to solar flares we presented here, it
is necessary to study the statistical
properties of an isolated UCS, i.e.\ to investigate the statistics
of the energy gain when an entire distribution of particle moves through
a UCS, all particles having random initial conditions.
This question belongs to the field of MHD in combination with
kinetic plasma physics (in what refers to anomalous resistivities).
Also needed is an understanding of
the spatial organization of an ensemble of co-existing UCSs and of their
connectivity and evolution. A first hint to how the UCS might be organized
spatially
comes from the cited inquiries of SOC models, which are in favor
of a global fractal structure with dimension around $1.8$.
The problem actually concerns the nature of 3-D, large scale,
magnetized MHD turbulence, and it involves theory as well as observations.

With the concrete specifications of the combined random walk to the
solar flare problem we made, we were able to
achieve HXR spectra which are compatible with the observations.
Important is that the model naturally leads to heating of the
plasma, or, more precise, it creates a heated population in the plasma.
This heated population can be expected to heat the entire background
plasma through collisional interactions on collisional time-scales,
and explaining in this way the observed delay
between the thermal soft X-ray and the non-thermal hard X-ray emission.
This mechanism of heat diffusion is though not included in our model.

%\subsection{Diffusion\label{Sec51}}

\noindent\textbf{Diffusion:}
Position- and velocity-space diffusion of the particles is anomalous,
the particles are highly super-diffusive in position space and
sub-diffusive in velocity-space (Fig.\ \ref{plotfour}).
The position-space super-diffusion implies that the particles
move fast through the active region. At one second, the particles have 
approximately moved a distance $3\,10^9\,{\rm cm}\approx L_{act}/3$,
so that fast particle-escape must be expected.
The particles may either diffuse to open field line regions,
from where they directly escape outwards, generating eventually type III 
radio bursts, or they may diffuse towards the lower atmosphere and give rise to
Bremsstrahlung emission. This fast time-scale of potential particle escape 
in the model is in qualitative accordance with the observed fast appearance 
of HXR emission and type III bursts in flares --- a conclusion which we can
though not draw too rigorously, since the free flight of the particles
we assume in the model is a simplification, an explicitly introduced
guiding magnetic field
topology might introduce effects which could to some degree alter this result.

%\subsection{Parametric study\label{Sec53}}
\noindent\textbf{Parametric study:}
We performed a extended parametric study of the model. The parameter
space is quite high dimensional, the results are dependent not just on the
power-law indices of the electric field and the spatial jump distributions,
but also on the respective ranges ($l_{min}$, $l_{max}$, $E_{min}$, $E_{max}$),
and finally on the mean value of the acceleration times (or acceleration 
lengths) and their standard
deviation. As a rule, the model yields power-law and also double power-law
tails
with varying power-law indices and varying ranges in the kinetic energies, and
also power-law shaped HXR spectra. The problem is to find parameter ranges
where the spectral index is compatible with the observations.

The power-law index of the spatial jump distribution we used here is
$1.2$, corresponding to the case of fractally distributed UCS with
fractal dimension $1.8$, and being thus consistent with
the results from SOC models (see Sec.\ \ref{Sec3}).

The power-law index of the electric field distribution
is $4.5$, a value we so far cannot justify with physical arguments. It is
anyway to note that the electric field used here is an effective one, and we
are not aware of a study we could compare it to. Choosing the lower limit
$E_{min}$ of the
electric fields as the Dreicer field is physically 
reasonable. Changes in
$E_{min}$ of the order of say $10$\% already influence the results quite
strongly, above all with respect to what highest energies are reached by the
particles. The setting of the maximum $E_{max}$ of the electric fields
to a relatively high value ($E_{max}=10^6\,E_D$) we justify with the following
argument: UCS are characterized not least by an intense current, which is very
likely to trigger some kinetic instabilities, with the result that the
resistivity is drastically increased by several orders of magnitude,
becoming anomalous, and, through Ohm's law, very strong electric
fields must be expected to appear in the short times until the energy
is dissipated.

The acceleration times are of the order of several milli-seconds, a time
scale which is roughly comparable to the fastest, non-thermal emissions
observed in flares, the so-called narrow-band, milli-second radio-spikes 
emission \citep[e.g.][]{Benz86}. The acceleration-lengths, on the other hand,
we do not consider to be directly comparable to observed phenomena, as 
explained in Sec.\ \ref{Sec31}.

Notably, we find cases where power-law tails are developed, i.e.\ where we
do find acceleration, the HXR emission is though too weak to
over-come the thermal emission. This finding is in favor of the
hypothesis that there is a big activity of small scale heating and
acceleration, which is though impossible to observe in the HXR since it
is covered by the thermal emission.

After all, we cannot claim that we have found the only and unique set
of parameter values for the problem under scrutiny, due to the
complexity of the model there may be other combinations of parameters
which give results compatible with the observations. It seems though
the values we chose are quite reasonable, physically.

A detailed parametric study, like the HXR spectral index as a function
of the electric field distribution power-law index, we plan to do in a future
study.

%\subsection{Ions}
\noindent\textbf{Ions:}
Application of the model (in the variant with 
prescribed acceleration-times) to ions yields results similar to the
case of the electrons, as long as the mean acceleration time is increased or 
stronger electric fields are applied.

In the first case of increased acceleration times, the process accelerating
the ions and the electrons is assumed to be the same, just that the ions
move with a lower velocity. Assuming that ions and electrons have the 
same temperature, the initial velocities of the protons are of the order of
$\sqrt{m_e/m_p}$ times smaller than the ones of the
electrons, where $m_e$ and $m_p$ are the electron and proton mass, 
respectively. The momentum increments in our model are independent of the 
masses, so that the corresponding increments in velocities are of the 
order of $m_e/m_p$ times smaller for the protons
than for the electrons. Assuming that the acceleration times are loosely
related to $1/v$, with $v$ a typical particle velocity, i.e.\ the particles
move along an internal trajectory in the UCS which is of similar length
for electrons and protons,
we find that the acceleration times must roughly be expected to be
between $\sqrt{m_p/m_e}\approx 42$ and $m_p/m_e\approx 1\,800$ times larger
for the protons than for the electrons. The $100$ times larger proton 
acceleration-times we applied in this article are thus in a justifiable 
range of values.

The advantage of the variant of the model with prescribed acceleration-lengths
is that without adjustment of the parameters from the electron case, the ions
get accelerated, though not to very high energies. The acceleration-lengths
variant has implicitly built-in the mass-dependence of the 
acceleration-efficiency.

%The second case of increased electric fields implies that the physical 
%process accelerating the
%ions and the one accelerating the electrons are not the same. The
%DC electric field which accelerates the electrons to the observed
%energies leaves the ions almost unaffected. The ions have thus to
%undergo their acceleration in a second process, which itself is
%ineffective on the electrons. This process most naturally could be
%shock acceleration by the shock waves which surround the UCS. This
%process is in principal identical to the one studied by \citet{Dec} 
% (see also Sec.\ \ref{Sec21}), with the only difference that the
%particles do not return to the same shock, but they meet different
%shock waves as they visit different UCS along their trajectories
% \citep[see also][]{Ana91}. 
%The mean electric field at a
%typical shock is $<E_s> \sim V_A B_0/c$, where $V_A$ is the local
%Alfv\'en speed,  $B_0$ the ambient magnetic field and $c$ the speed
%of light. For the parameters introduced in Sec.\ \ref{Sec4}, we 
%find $<E_s>\approx 0.7\,$statvolt/cm ($2.1\,10^4\,$V/m). On the other
%hand, the Dreicer electric field used in the electron acceleration
%is $E_D\approx 6.6\,10^{-7}\,$ ($2.0\,10^{-2}\,$V/m),
%so that $<E_s>$ is roughly $10^6$ times larger than $E_D$, and the
%application of a larger electric field to the ions is justified.

%\subsection{Comparison to observations}
\noindent\textbf{Comparison to observations:}
Benz and his collaborators have
repeatedly stressed and discussed in detail the
observational fact of 
fragmentation of the energy release \citep[e.g.][]{Benz94}.
They give evidence that the coronal acceleration region is probably co-spatial 
with the decimetric spikes emitting volume \citep{Benz03},
which in turn is highly fragmented in space and time.  
Another observational result
favoring the fragmentation of the acceleration region is the
appearance of groups of type III bursts during the impulsive phase
of flares \citep{Benz94}.
Our model incorporates the scenario of fragmented energy release,
it is actually one of its basic assumptions.

The observed non-thermal HXR spectra are characterized by a
near power-law form (often double power-law), with spectral index
which varies in the course of a flare, which is though always
clearly above $2$, and which is anti-correlated with the HXR flux
\citep[e.g.][]{Hudson02,Krucker02,Lin02,Saint02,Sui02}. 
According to \citet{Lin03}, the spectra up to $400\,$keV are generated
purely by the electrons, from $400\,$keV to $1.4$MeV electron Bremsstrahlung
dominates, ion emission starts though increasingly to alter the spectra, and
above $1.4\,$MeV the spectra are dominated by the ion line-emission.
The HXR spectra we presented are generated by electron Bremsstrahlung, 
they extend up to $500\,$keV, and from their power-law shape and spectral 
index they can be considered compatible with the observations. 
Slight variations in the model parameters will be reflected in variations 
of the spectral index. In particular, the variant of the model in which the 
acceleration-lengths are prescribed has a tendency to yield double power-laws
for larger times, it seems that the most energetic particles get less and less
increase in energy as they become faster, leading to a steepening of the 
power-law tail.
The double power-law becomes though too flat at low energies for too 
large times --- actually there the model in its current form reaches its 
limitations, it is constructed only for the very first period of the 
impulsive phase, since collisional losses and losses of particles out of the 
active region are not included.

The observed microwave emission seems not to follow a very
clear, universal spectral shape. Usually, the spectrum increases
until a frequency of a few hundred MHz, and then it falls off quite
steeply, sometimes in a clear power-law form, sometimes more in an
exponential manner \citep[see e.g.][]{Lee03,Flei03}.
We can thus conclude that the microwave spectrum yielded by our model
seems compatible with the observations. It is to note that the 
microwave spectra 
and the degree of polarization depend quite sensitively on the assumed 
background magnetic field strength $B_0$. While the shape of the spectrum 
remains
qualitatively the same, the steepnesses of the increase at low frequencies
and of the decrease at high frequencies quantitatively change, 
their closeness to power-law shapes varies,
and 
the turn-over frequency and the relative intensity of the o- and x-mode
change,
if the value of $B_0$ changes.

\noindent\textbf{On the number problem:}
The number problem refers to the question whether an acceleration
mechanism is efficient and fast enough to generate the large number of 
accelerated particles as they are inferred from the observations.
\citet[and references therein]{Mill97} mention that roughly
$10^{34}$ to $10^{37}\,$electrons/sec are accelerated --- the number
cannot be very precise, it is indirectly derived 
and depends on several assumptions on emission
mechanisms and on mostly assumed active region parameters.

In order to check the efficiency of our model, we first determine
numerically the total energy $E_{kin,t}$ of the $n_e$ electrons which are in 
the power-law tail, with $n_e$ now a free parameter,
$E_{kin,t}=n_e \int_{5\,{\rm keV}}^{E_{max}} p(E_{kin},t=1\,{\rm sec}) 
E_{kin}\,dE_{kin}$, where $p(E_{kin},t=1\,{\rm sec})$ is the 
numerically given kinetic energy distribution at $t=1\,$sec, as yielded 
by our model, and the tail is defined to start above $5\,$keV 
(see Sec.\ \ref{Sec42} and Fig.\ \ref{plotfive}).
Integrating numerically, we find
$E_{kin,t}=n_e\,9\,10^{-9}\,$erg. If we assume a large, but not huge, flare
in which say $10^{30}\,$erg of energy are released during say $100\,$sec,
we find that $n_e\approx 10^{36}\,$electrons/sec must be accelerated.
In a typical coronal active region of volume 
$L_{acc}^3=10^{30}\,$cm$^3$ and with particle density  
$n_0=10^{10}\,$cm$^{-3}$, there is a total number of $10^{40}$ 
particles . During the flare of $100\,$sec duration,
$10^{38}$ particles are accelerated in total and potentially leave,
corresponding to $1\%$ of the initial particles.
With these numbers, it seems that there is no essential deplenishment
of the flaring volume, and there is thus no need for a secondary
mechanism of replenishment. We just note that the volume assumed in this 
estimate is that of the entire active region since the high diffusivity 
of the particles (see earlier in this section) and the nature of SOC 
(see Fig.\ \ref{plotone}) let us expect that indeed the entire active
region contributes to the acceleration of particles in flares.  

Doing the same kind of analysis for the ions, with the difference
that the power-law tail starts now above $10\,$keV (see Fig.\ \ref{ploteight}),
we find a total energy of the accelerated protons 
$E_{kin,t}^{(i)}=n_e\,1.0\,10^{-8}\,$erg (in the case of increased 
acceleration times, see Sec.\ \ref{ions}) or
$E_{kin,t}^{(i)}=n_e\,2.5\,10^{-8}\,$erg (in the case of increased 
electric fields), which is roughly $1.1$ to $3$ times
larger than the electron energy in the tail, so that less protons
have to be accelerated to account for the observed energies.
As a remark, we note that this result does not imply an imbalance between
electron- and proton-acceleration, there may well exist
an unobserved, 'dark' population of particles, above all protons at low
energies, so that the numbers of accelerated electrons and protons may 
actually be equal.

%We can thus claim that our model is efficient enough 
%and yields numbers of accelerated particles which are reasonable and 
%compatible with the ones derived from observations. 
%We do though not dare to claim on the same firm grounds that there 
%is indeed no need for
%replenishment, since the numbers are too vague and too many
%loose assumptions on coronal parameters had to be made.
%In favor of the non-need of replenishment is the high diffusivity 
%of particles, which allows a large mobility of the particles through
%the entire flaring region --- unless effects of an explicitly introduced
%guiding magnetic topology would alter the high diffusivity (see the 
%discussion in Sec.\ \ref{Sec51}).

\subsection{Main achievements of our model and open questions}

The main results and characteristics of our model are:
(1) The acceleration process is extremely fast. Electrons reach
  MeV-energies in $0.01\,$s, and ions approach $10\,$MeV in $1\,$s.
(2) Our model exhibits strong super-diffusion of particles in
  position-space. One important effect of the fast transport of accelerated
  particles is that they can be expected to reach fast open field-line regions.
  Accelerated particles will thus easily
  escape into the upper corona and the interplanetary space if 
  closed and open field line regions co-exist inside a large active
  region.
(3) The energy spectra formed for both species are
  Maxwellians with steep power-law tails.
(4) The HXR spectra are compatible with the observations (power-law shapes,
   spectral indices above 2).
(5) The micro-wave spectra are qualitatively compatible with the 
  observations.
(6) We observe efficient, fast plasma heating.
  Energy release, acceleration, and heating are unified,
  they are the result of the same process.
(7) The results for electrons and ions are quite similar,
  the ions need though either longer acceleration times or 
  stronger electric fields. 
(8) The total number of accelerated particles is compatible 
  with numbers derived from observations.
(9) Electrons and ions reach their maximum energy on a
  very fast time scale, the system adjusts itself quickly to the
  conditions imposed by the magnetic structure and the energy
  release process (in the model through the modification of the chosen
  probabilities $P_1(s)$, $P_2(E)$, $P_3(\tau)$, see Sec.\ \ref{Sec31}). 
  In other words, the acceleration would follow the
  evolution of $P_1(s,t)$ and $P_2(E,t)$ very closely, it instantaneously
  adjusts itself to the instantaneous coronal conditions.

%\subsection{Open questions}

We have shown that radiative losses due to synchrotron radiation
are negligible for the particle dynamics, due to
the energies and the densities considered
here, but in general they have to be
incorporated in the evolution of particles, especially if this
process is applied to astrophysical sources.

Not included in our study are the collisional losses
of the particles during their free travels. They possibly
would affect mainly the particles with low velocities, a fraction
of the particles might be thermalized during their free flights.
Also not included are the radiation associated with collisions,
the possible thin target HXR emission of the system, and the escape 
of particles, i.e.\ the diffusive loss of particles out of the active region. 

It is also to be mentioned that in the random walk approach we
made here, the spatial evolution of the particles is treated in a
simplified way, we have included the magnetic
topology only implicitly through the distribution $P_1(s)$
of spatial jumps. An explicitly included topology would cause effects
like escape, mirroring, trapping, which might affect the timing
and the spatial structure of a flare to some degree (e.g.\ 'loop-tops',
'loop-top with a helmet', 'foot-points', etc.).

\section{Conclusion}

In this article, we  shifted the emphasis of the acceleration
process in active regions away from the details of 
specific mechanism(s) involved and focused on the global aspects of the 
active region,
its evolution in space and time, and on the stochastic 
nature of the acceleration process. Our basic assumptions are: 
(1) acceleration is a local process in and near the UCS, (2) the
UCS are distributed in a complex way inside the large-scale 3-D magnetic 
topology of the active region, and (3) acceleration is the result of 
multiple interactions with different UCS.
These assumptions can be summarized in the statement that  
energy dissipation and particle acceleration in flares are fragmented.
We
achieved for the first time to connect the accelerator to the
3-D magnetic topology and the energy
release process. 
From our results,
we can safely conclude that the complexity of the 3-D magnetic
field topology in active regions, in combination with the stresses imposed by 
the convection zone onto it,
forms a highly efficient accelerator.

The adequate tool for modeling the stochastic nature of particle 
acceleration in flares 
is continuous time random walk in position- and velocity-space. 
This new approach opens up the way for the understanding 
of a variety of acceleration phenomena, e.g.\ acceleration before and
after a flare, acceleration without a flare, long lasting
acceleration etc.

We feel that the road opened in this article is new and still
unexplored in astrophysics. As mentioned,
it can provide answers to a number of open problems
which have remained unsolved when it was attempted to use one UCS
(above or inside a flaring loop) and one (uncorrelated with the
energy release) acceleration mechanism to explain phenomena of
acceleration. Many new questions are
still open and we plan to return to these issues in a forth-coming
article.

\acknowledgments{This work was supported in part by the Research
Training Network (RTN) 'Theory, Observation and Simulation of Turbulence 
in Space Plasmas', funded by the European Commission (contract No.\ 
HPRN-eT-2001-00310). The authors would like to thank A.\ Anastasiadis
for helpful discussions and comments on the manuscript, and the referee
for his constructive critique, which helped to improve the article.
T.\ Fragos, M.\ Rantziou and I.\ Sandberg are acknowledged for their
help in image processing.
L. V. was a member of the team 'Critical Problems in Solar Flare
Physics', which met three times during the period from 2000 to 2003. 
He is most
grateful to the International Space Science Institute (ISSI) in Bern for
supporting his participation. He is also grateful to the team members P.
Cargill (team leader), Arnold Benz, Hugh Hudson, Brigitte Schmieder,
George Simnett for the stimulating discussions and constructive
suggestions.}

\end{document}